\documentclass[fleqn,usenatbib,useAMS,onecolumn]{mnras}

\usepackage{graphicx}	
\usepackage{amsmath}	
\usepackage{nccmath}
\usepackage{amssymb}	

\usepackage[dvipsnames]{xcolor}  
\usepackage{soul}
\usepackage{xcolor}
\usepackage{subfig}
\usepackage{floatrow}
\usepackage{color}
\usepackage{ulem}
\usepackage{comment}
\usepackage{bm}
\usepackage{cuted}
\usepackage{flushend}
\usepackage{cancel}

\usepackage{macros_vdb} 

\begin{document}


\title[Impulsive Encounters]
      {A Fully General, Non-Perturbative Treatment of Impulsive Heating}

\author[Banik and van den Bosch]{%
   Uddipan Banik$^{1}$\thanks{E-mail: uddipan.banik@yale.edu},
   Frank~C.~van den Bosch$^1$
\vspace*{8pt}
\\
   $^1$Department of Astronomy, Yale University, PO. Box 208101, New Haven, CT 06520, USA
   }


\date{}

\pagerange{\pageref{firstpage}--\pageref{lastpage}}
\label{firstpage}
\pubyear{2020}

\maketitle


\begin{abstract}
  Impulsive encounters between astrophysical objects are usually treated using the distant tide approximation (DTA) for which the impact parameter, $b$, is assumed to be significantly larger than the characteristic radii of the subject, $r_\rmS$, and the perturber, $r_\rmP$. The perturber potential is then expanded as a multipole series and truncated at the quadrupole term. When the perturber is more extended than the subject, this standard approach can be extended to the case where $r_\rmS \ll b < r_\rmP$. However, for encounters with $b$ of order $r_\rmS$ or smaller, the DTA typically overpredicts the impulse, $\Delta \bv$, and hence the internal energy change of the subject, $\Delta E_{\rm int}$. This is unfortunate, as these close encounters are the most interesting, potentially leading to tidal capture, mass stripping, or tidal disruption. Another drawback of the DTA is that $\Delta E_{\rm int}$ is proportional to the moment of inertia, which diverges unless the subject is truncated or has a density profile that falls off faster than $r^{-5}$. To overcome these shortcomings, this paper presents a fully general, non-perturbative treatment of impulsive encounters which is valid for any impact parameter, and not hampered by divergence issues, thereby negating the necessity to truncate the subject. We present analytical expressions for $\Delta \bv$ for a variety of perturber profiles, apply our formalism to both straight-path encounters and eccentric orbits, and discuss the mass loss due to tidal shocks in gravitational encounters between equal mass galaxies.
\end{abstract}




\section{Introduction}

When an extended object, hereafter the subject, has a gravitational encounter with another massive body, hereafter the perturber, it induces a tidal distortion that causes a transfer of orbital energy to internal energy of the body (i.e., coherent bulk motion is transferred into random motion).  Gravitational encounters therefore are a means by which two unbound objects can become bound (`tidal capture'), and ultimately merge. They also cause a heating and deformation of the subject, which can result in mass loss and even a complete disruption of the subject.  Gravitational encounters thus play an important role in many areas of astrophysics, including, among others, the merging of galaxies and dark matter halos \citep[e.g.,][]{Richstone.75, Richstone.76, White.78, Makino.Hut.97, Mamon.92, Mamon.00}, the tidal stripping, heating and harassment of subhalos, satellite galaxies and globular clusters \citep[e.g.,][]{Moore.etal.96b, Gnedin.etal.99, vdBosch.etal.18a,DuttaChowdhury.etal.20}, the heating of discs \citep[][]{Ostriker.etal.72}, the formation of stellar binaries by two-body tidal capture \citep[][]{Fabian.etal.75, Press.Teukolsky.77, Lee.Ostriker.86}, and the disruption of star clusters and stellar binaries \citep[e.g.,][]{Spitzer.58, Heggie.75, Bahcall.etal.85}. Throughout this paper, for brevity we will refer to the constituent particles of the subject as `stars'.
\begin{figure*}
\includegraphics[width=0.9\textwidth]{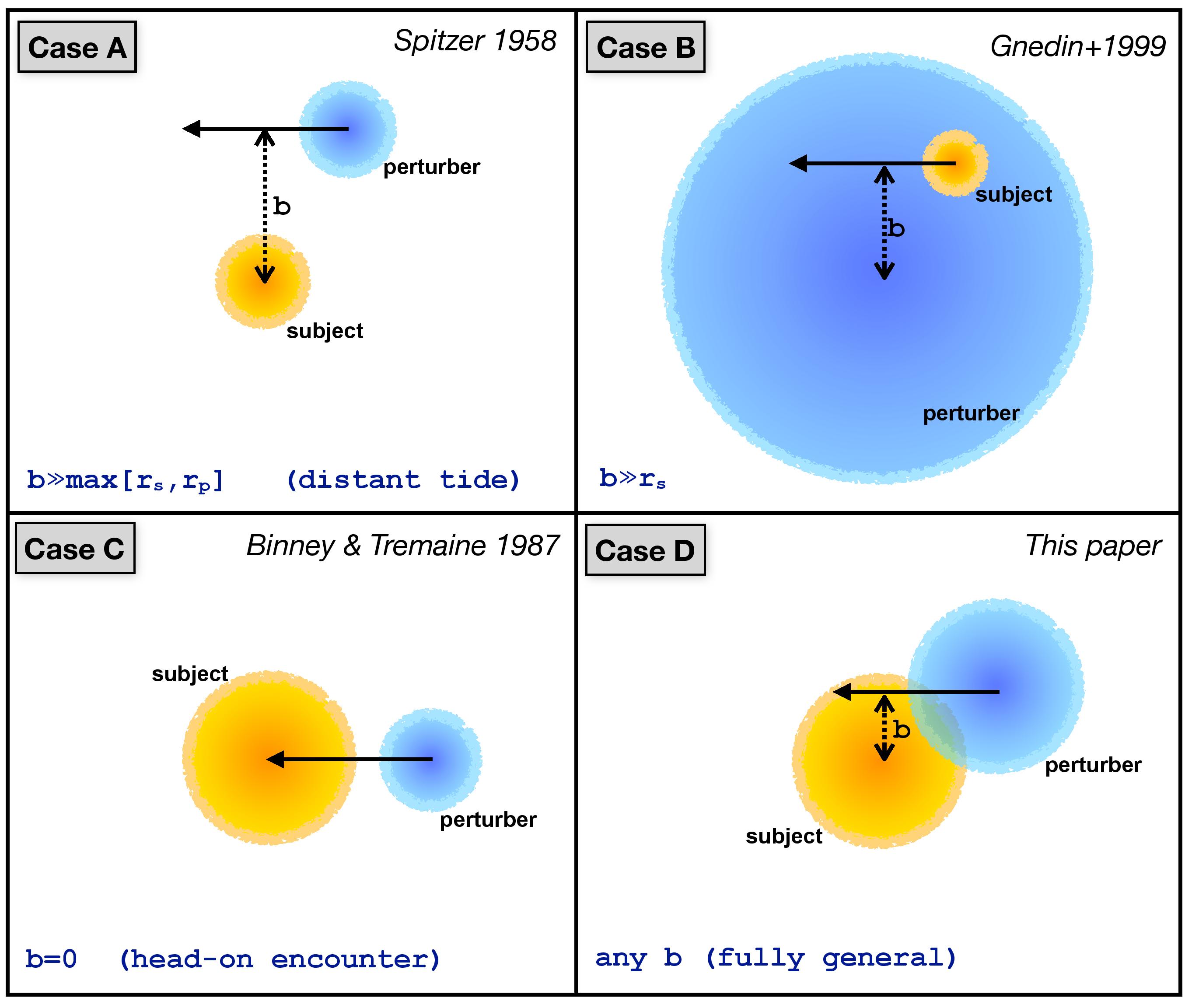}
   \caption{A pictorial comparison of impulsive encounters $(\vp\gg \sigma)$ under certain conditions for the impact parameter $b$. In the upper-right corner of each panel we cite the paper in which the impulsive energy transfer for this case was first worked out. This paper presents the fully general case D (no constraint on $b$), as depicted in the lower right-hand panel.}
\label{fig:schematic_approximations}
\end{figure*}
A fully general treatment of gravitational encounters is extremely complicated, which is why they are often studied using numerical simulations. However, in the impulsive limit, when the encounter velocity is large compared to the characteristic internal velocities of the subject, the encounter can be treated analytically. In particular, in this case, one can ignore the internal motion within the subject (i.e., ignore the displacements of the stars during the encounter), and simply compute the velocity change (the impulse) of a star using
\begin{equation}
\Delta\bv = -\int \nabla \Phi_\rmP \, \rmd t\,,
\end{equation}
where $\Phi_\rmP$ is the potential due to the perturber. And since the encounter speed, $\vp$, is high, one can further simplify matters by considering the perturber to be on a straight-line orbit with constant speed.

The impulse increases the specific kinetic energy of the subject stars by
\begin{equation}\label{dEstar}
\Delta \varepsilon = \bv \cdot \Delta \bv + {1 \over 2} (\Delta v)^2\,.
\end{equation}
Since the potential energy of the stars remains invariant during (but not after) the impulse, the increase in total {\it internal} energy of the subject is given by
\begin{equation}
\Delta E_{\rm int} = \int \rho_\rmS(\br) \Delta \varepsilon(\br) \, \rmd^3\br - {1 \over 2} M_\rmS (\Delta v_{\rm CM})^2\,.
\end{equation}
Here $M_\rmS$ and $\rho_\rmS(\br)$ are the mass and density profile of the subject and $\Delta \bv_{\rm CM}$ is the velocity impulse of the centre-of-mass of the subject.

If the encounter, in addition to being impulsive, is also distant, such that the impact parameter $b$ is much larger than the scale radii of the subject ($r_\rmS$) and the perturber ($r_\rmP$), i.e., $b \gg \max(r_\rmS, r_\rmP)$, then the internal structure of the perturber can be ignored (it can be treated as a point mass), and its potential can be expanded as a multipole series and truncated at the quadrupole term. This `distant tide approximation' (hereafter DTA, depicted as case A in Fig.~\ref{fig:schematic_approximations}) was first used by \citet[][hereafter S58]{Spitzer.58} to study the disruption of star clusters by passing interstellar clouds.  In particular, Spitzer showed that, for a spherical subject mass, $M_\rmS$, an impulsive encounter results in an internal energy increase
\begin{equation}\label{dESpitzer}
\Delta E_{\rm int} = {4 M_\rmS \over 3} \, \left({G \, M_\rmP \over \vp} \right)^2 \, {\langle r^2 \rangle \over b^4}\,,
\end{equation}
with
\begin{equation}\label{r2aver}
\langle r^2 \rangle = \frac{4 \pi}{M_\rmS} \int \rho_\rmS(r) \, r^4 \, \rmd r
\end{equation}
(see also Table~\ref{tab:comparison}). Note that $\Delta E \propto b^{-4}$, indicating that closer encounters are far more efficient in transferring energy than distant encounters.  However, as shown by \cite{Aguilar.White.85} using numerical simulations, equation~(\ref{dESpitzer}) is only accurate for relatively large impact parameters, $b \gta 10\max(r_\rmS, r_\rmP)$, for which $\Delta E_{\rm int}$ is typically extremely small (and thus less interesting).

This situation was improved upon by \citet[][hereafter GHO99]{Gnedin.etal.99}, who modified the treatment by S58 so that it can also be used in cases where $r_\rmS \ll b < r_\rmP$ (see case B in Fig.~\ref{fig:schematic_approximations}). This describes circumstances in which the subject is moving inside the perturber potential (i.e., a globular cluster moving inside a galaxy, or a satellite galaxy orbiting the halo of the Milky Way). As shown by GHO99, the resulting $\Delta E_{\rm int}$ in this case is identical to that of equation~(\ref{dESpitzer}) but multiplied by a function $\chi_{\rm st}(b)$, that depends on the detailed density profile of the perturber (see  Table~\ref{tab:comparison}).
\begin{table*}
\centering
\begin{tabular}{lllll}
 \hline
 Case & Impact parameter & $\Delta E_{\rm int}$ & Source\\
 (1) & (2) & (3) & (4)\\
 \hline \\
A & \begin{minipage}{2cm}\begin{equation}b\gg \max{\left(r_\rmS,r_\rmP\right)}\nonumber \end{equation}\end{minipage}    & \begin{minipage}{2 cm}\begin{equation}\frac{4 M_\rmS}{3}{\left(\frac{GM_\rmP}{\vp}\right)}^2 \frac{\left<r^2\right>}{b^4},\nonumber\end{equation}\end{minipage} & [A] \\ \\
& & \begin{minipage}{2cm}\begin{align}\left<r^2\right>=\frac{4\pi}{M_\rmS}\int_0^{r_{\rm trunc}}\rmd r\, r^4\rho_\rmS(r)\nonumber\end{align}\end{minipage} &
\\ \\ \\ \\
B      & \begin{minipage}{2cm}\begin{equation}b\gg r_\rmS\nonumber \end{equation}\end{minipage}    & \begin{minipage}{2cm}\begin{equation}\frac{4 M_\rmS}{3}{\left(\frac{GM_\rmP}{\vp}\right)}^2 \left<r^2\right>\frac{\chi_{\rm st}(b)}{b^4},\nonumber \end{equation}\end{minipage} & [B]\\ \\
& & \begin{minipage}{2cm}\begin{equation}\chi_{\rm st}=\frac{1}{2}\left[{\left(3J_0-J_1-I_0\right)}^2+{\left(2I_0-I_1-3J_0+J_1\right)}^2+I^2_0\right],\nonumber\end{equation}\end{minipage}\\ \\
& & \begin{minipage}{2cm}\begin{equation}I_k(b)=\int_1^{\infty}\mu_k(b\zeta)\frac{\rmd \zeta}{\zeta^2{\left(\zeta^2-1\right)}^{1/2}},\nonumber\end{equation}\end{minipage}\\ \\
& & \begin{minipage}{2cm}\begin{equation}J_k(b)=\int_1^{\infty}\mu_k(b\zeta)\frac{\rmd \zeta}{\zeta^4{\left(\zeta^2-1\right)}^{1/2}} \;\;(k=0,1),\nonumber\end{equation}\end{minipage} \\ \\
& & \begin{minipage}{2cm}\begin{equation}\mu_0(R)=\frac{M_\rmP(R)}{M_\rmP}, \;\;\mu_1(R)=\frac{\rmd \mu_0(R)}{\rmd \ln{R}}\nonumber\end{equation}\end{minipage}\\ \\ \\ \\
C   & \begin{minipage}{2cm}\begin{equation}b=0\nonumber\end{equation}\end{minipage}     & \begin{minipage}{2cm}\begin{equation}4\pi{\left(\frac{GM_\rmP}{\vp}\right)}^2 \int_0^{r_{\rm trunc}} \frac{\rmd R}{R}I^2_0(R)\Sigma_\rmS(R),\nonumber\end{equation}\end{minipage} & [C] \\ \\
& & \begin{minipage}{2cm}\begin{equation}\Sigma_\rmS(R)=2\int_R^{r_{\rm trunc}}\rho_\rmS(r)\frac{r\,\rmd r}{\sqrt{r^2-R^2}}\nonumber\end{equation}\end{minipage} \\ \\ \\ \\
D &  Any $b$ & \begin{minipage}{2cm}\begin{equation}2{\left(\frac{GM_\rmP}{\vp}\right)}^2 \left[\int_0^{\infty}\rmd r\,r^2\rho_\rmS(r)\calJ(r,b)-\calV(b)\right],\nonumber\end{equation}\end{minipage} & [D]\\ \\
& & \begin{minipage}{2cm}\begin{align}\calJ(r,b)=\int_0^{\pi}\rmd \theta \sin{\theta} \int_0^{2\pi}\rmd \phi\;s^2 I^2(s),\;\;\; s^2=r^2\sin^2{\theta}+b^2-2br\sin{\theta}\sin{\phi},\nonumber\end{align}\end{minipage} \\ \\
& & \begin{minipage}{2cm}\begin{equation}\calV(b)=\frac{1}{M_\rmS}{\left[\int_0^{\infty}\rmd r\, r^2 \rho_\rmS(r)\calJ_{\rm CM}(r,b)\right]}^2,\nonumber\end{equation}\end{minipage}\\ \\
& & \begin{minipage}{2cm}\begin{equation}\calJ_{\rm CM}(r,b)=\int_0^{\pi}\rmd \theta \sin{\theta} \int_0^{2\pi}\rmd \phi\; I(s)\,\left[b-r\sin{\theta}\sin{\phi}\right],\nonumber\end{equation}\end{minipage}\\ \\
& & \begin{minipage}{2cm}\begin{equation}I(s) = \int_0^{\infty} \rmd \zeta\, \frac{1}{R_\rmP}\frac{\rmd\Tilde{\Phi}_\rmP}{\rmd R_\rmP},\nonumber\end{equation}\end{minipage}\\ \\
& & \begin{minipage}{2cm}\begin{equation}\Tilde{\Phi}_\rmP = \Phi_\rmP/(GM_\rmP),\; R_\rmP=\sqrt{s^2+\zeta^2}\nonumber\end{equation}\end{minipage} \\ \\
 \hline
\end{tabular}
\caption{Full set of expressions needed to compute $\Delta E_{\rm int}$ (considering an impulsive encounter along a straight-line orbit) for the four cases depicted in Fig.~\ref{fig:schematic_approximations}. Column [2] lists the range of impact parameters for which these expressions are accurate, and column [4] lists the relevant reference, where [A],[B],[C] and [D] correspond to \protect\cite{Spitzer.58}, \protect\cite{Gnedin.etal.99}, \protect\cite{vdBosch.etal.18a}, and this paper, respectively.}
\label{tab:comparison}
\end{table*}

Although this modification by GHO99 significantly extends the range of applicability of the impulse approximation, it is still based on the DTA, which requires that $b \gg r_\rmS$.  For smaller impact parameters, $\Delta E_{\rm int}$ computed using the method of GHO99 can significantly overpredict the amount of impulsive heating (see \S\ref{sec:plummer_straight_orbit}). There is one special case, though, for which $\Delta E_{\rm int}$ can be computed analytically, which is that of a head-on encounter ($b=0$; see Case C in Fig.~\ref{fig:schematic_approximations}) when both the perturber and the subject are spherical.  In that case, as shown in \cite{Binney.Tremaine.87}, the symmetry of the problem allows a simple analytical calculation of $\Delta E_{\rm int}$ (see  Table~\ref{tab:comparison}). This was used by \citet{vdBosch.etal.18a} to argue that one may approximate $\Delta E_{\rm int}(b)$ for {\it any} impact parameter, $b$, by simply setting $\Delta E_{\rm int}(b) = \min[\Delta E_{\rm dt}(b), \Delta E_0]$.  Here $\Delta E_{\rm dt}(b)$ is the $\Delta E_{\rm int}(b)$ computed using the DTA of GHO99 (case B in  Table~\ref{tab:comparison}), and $\Delta E_0$ is the $\Delta E_{\rm int}$ for a head-on encounter (case C in  Table~\ref{tab:comparison}). Although a reasonable assumption, this approach is least accurate exactly for those impact parameters ($b \sim r_\rmS$) that statistically are expected to be most relevant\footnote{For a uniform background of perturbers, the probability that an encounter has an impact parameter in the range $b$ to $b+\rmd b$ is $P(b) \rmd b \propto b \rmd b$, such that the total $\Delta E$ due to many encounters is dominated by those with $b \sim r_\rmS$.}.

Another shortcoming of using the DTA is that $\Delta E_{\rm int}$ is found to be proportional to $\langle r^2 \rangle$, the mean squared radius of the subject (see equation~[\ref{r2aver}] and Table~1). For most density profiles typically used to model galaxies, dark matter halos, or star clusters, $\langle r^2 \rangle$ diverges, unless the asymptotic radial fall-off of the density is steeper than $r^{-5}$, or the subject is physically truncated. Although in reality all subjects are indeed truncated by an external tidal field, it is common practice to truncate the density profile of the subject at some arbitrary radius rather than a physically motivated radius. And since $\langle r^2 \rangle$ depends strongly on the truncation radius adopted (see \S\ref{sec:plummer_straight_orbit}), this can introduce large uncertainties in the amount of orbital energy transferred to internal energy during the encounter.

In this paper, we develop a fully general, non-perturbative formalism to compute the internal energy change of a subject due to an impulsive encounter. Unlike in the DTA, we do not expand the perturber potential as a multipole series, which assures that our formalism is valid for any impact parameter. For the impulse approximation to be valid, the encounter time $\tau=b/\vp$ has to be small compared to the typical orbital timescale of the subject stars. However, in the distant tide limit, when $b$ is large, the encounter time will also typically be large, rendering the impulse approximation invalid unless $\vp$ is very large. In other words, although there are cases for which the DTA and the impulse approximation are both valid, often they are mutually exclusive.  Our formalism, being applicable to all impact parameters, is not hampered by this shortcoming. Moreover, our expression for the internal energy change does not suffer from the $\langle r^2 \rangle$ divergence issue mentioned above, but instead converges, even for infinitely extended systems. This alleviates the problem of having to truncate the galaxy at an arbitrary radius.

This paper is organized as follows. In \S\ref{sec:straight_orbit}, we present our general formalism to compute the impulse and the energy transferred in impulsive encounters along straight-line orbits. In \S\ref{sec:special} we apply our formalism to several specific perturber density profiles. In \S\ref{sec:eccentric_orbit} we further generalize the formalism to encounters along eccentric orbits, incorporating an adiabatic correction \citep[][]{Gnedin.Ostriker.99} to account for the fact that for some subject stars, those with short dynamical times, the impact of the encounter is adiabatic rather than impulsive. In \S\ref{sec:mass_loss}, as an astrophysical application of our formalism, we discuss the mass loss of \citet{Hernquist.90} spheres due to tidal shocks during mutual encounters. Finally we summarise our findings in \S\ref{sec:conclusion}.

\section{Encounters along straight-line orbits}
\label{sec:straight_orbit}

\begin{figure}
\includegraphics[width=0.75\textwidth]{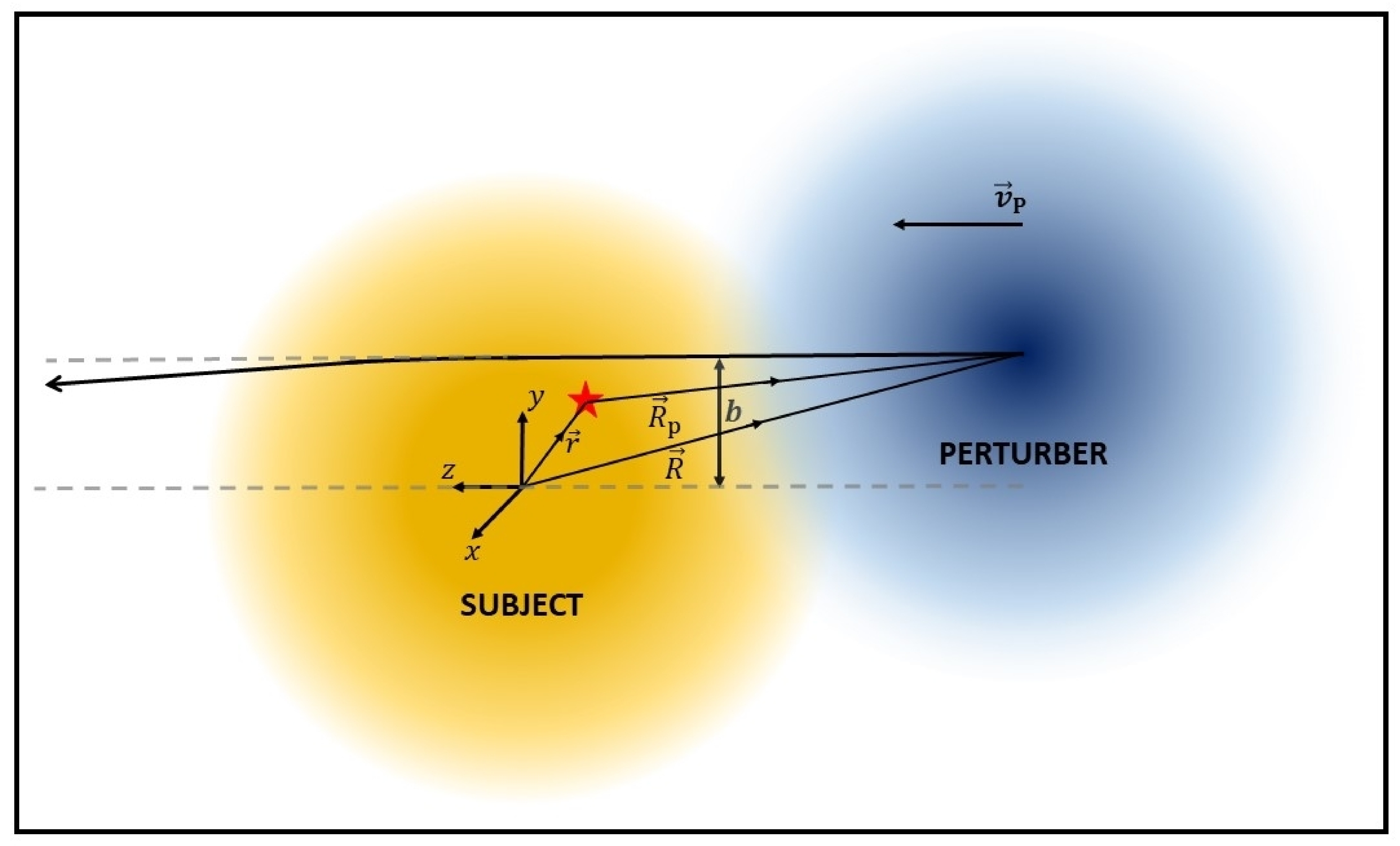}
   \caption{Illustration of the geometry of an impulsive encounter along a nearly straight orbit, specifying the coordinate axes and radial vectors used throughout this paper.}
\label{fig:schematic_straight_orbit}
\end{figure}

Consider the gravitational encounter between two self-gravitating bodies, hereafter `galaxies'. In this section we assume that the two galaxies are mutually unbound to begin with and approach each other along a hyperbolic orbit with initial, relative velocity $\vp$ and impact parameter $b$. For sufficiently fast encounters (large $\vp$), the deflection of the galaxies from their original orbits due to their mutual gravitational interaction is small and we can approximate the orbits as a straight line. We study the impulsive heating of one of the galaxies (the subject) by the gravitational field of the other (the perturber). Throughout this paper we always assume the perturber to be infinitely extended, while the subject is either truncated or infinitely extended. For simplicity we consider both the perturber and the subject to be spherically symmetric, with density profiles $\rho_\rmP(r)$ and $\rho_\rmS(r)$, respectively. The masses of the subject and the perturber are denoted by $M_\rmS$ and $M_\rmP$ respectively, and $r_\rmS$ and $r_\rmP$ are their scale radii. We take the centre of the unperturbed subject as the origin and define $\hat{\bz}$ to be oriented along the relative velocity $\bvp$, and $\hat{\by}$ perpendicular to $\hat{\bz}$ and directed towards the orbit of the perturber. The position vector of a star belonging to the subject is given by $\br$, that of the COM of the perturber is $\bR$ and that of the COM of the perturber with respect to the star is $\bR_\rmP=\bR-\br$ (see Fig.~\ref{fig:schematic_straight_orbit}).
\subsection{Velocity perturbation up to all orders}
\label{sec:velocity_straight_orbit}

During the encounter, the perturber exerts an external, gravitational force on each subject star. The potential due to the perturber flying by with an impact parameter $b$, on a particle located at $\br = (x,y,z)$ is a function of the distance to the particle from its center, $R_\rmP = \left|\bR-\br\right| = \sqrt{x^2+{\left(b-y\right)}^2+{\left(z-\vp t\right)}^2}$. The acceleration of the star due to the perturbing force is directed along $\widehat{\bR}_\rmP = \widehat{\bR} - \hat{\br} = \left[-x\hat{\bx}+\left(b-y\right)\hat{\by}-\left(z-\vp t\right)\hat{\bz}\right]/R_\rmP$, and is equal to
\begin{align}
&\ba_\rmP = -\nabla \Phi_\rmP = \frac{1}{R_\rmP}\frac{\rmd\Phi_\rmP}{\rmd R_\rmP} \left[-x\hat{\bx}+\left(b-y\right)\hat{\by}-\left(z-\vp t\right)\hat{\bz}\right]. 
\label{force}
\end{align}
We assume that the perturber moves along a straight-line orbit from $t\rightarrow -\infty$ to $t\rightarrow \infty$. Therefore, under the perturbing force, the particle undergoes a velocity change, 
\begin{align}
&\Delta \bv = \int_{-\infty}^{\infty} \rmd t\, \ba_\rmP = \int_{-\infty}^{\infty} \rmd t\,\frac{1}{R_\rmP}\frac{\rmd\Phi_\rmP}{\rmd R_\rmP} \left[-x\hat{\bx}+\left(b-y\right)\hat{\by}-\left(z-\vp t\right)\hat{\bz}\right].
\label{deltav0}
\end{align}
The integral along $\hat{\bz}$ vanishes since the integrand is an odd function of $\left(z-\vp t\right)$. Therefore the net velocity change of the particle occurs along the $x-y$ plane and is given by 
\begin{align}
\Delta \bv = \frac{2 G M_\rmP}{\vp} I(s) \left[-x\hat{\bx} + (b-y)\hat{\by}\right],
\label{deltav1}
\end{align}
where $s^2=x^2+{\left(b-y\right)}^2$. The integral $I(s)$ is given by
\begin{align}
I(s) = \int_0^{\infty} \rmd \zeta\, \frac{1}{R_\rmP}\frac{\rmd\Tilde{\Phi}_\rmP}{\rmd R_\rmP}\,.
\label{I}
\end{align}
Here $\Tilde{\Phi}_\rmP=\Phi_\rmP/(GM_\rmP)$, $R_\rmP=\sqrt{s^2+\zeta^2}$, and $\zeta = \vp t-z$. Note that the above expression for $\Delta \bv$ is a slightly modified version of that obtained by \citet[][equation (3) of their paper]{Aguilar.White.85}. The integral $I(s)$ contains information about the impact parameter of the encounter as well as the detailed density profile of the perturber. Table.~\ref{tab:Is} lists analytical expressions for a number of different perturber potentials, including a point mass, a \cite{Plummer.11} sphere, a \cite{Hernquist.90} sphere, a NFW profile \citep[][]{Navarro.etal.97}, the Isochrone potential \citep[][]{Henon.59, Binney.14}, and a Gaussian potential. The latter is useful since realistic potentials can often be accurately represented using a multi-Gaussian expansion \citep[e.g.][]{Emsellem.etal.94, Cappellari.02}.
\begin{table*}
\centering
\begin{tabular}{lllll}
 \hline
 Perturber profile & $\Phi_\rmP(r)$ & $I(s)$ \\
 (1) & (2) & (3) \\
 \hline 
 \\
Point mass      & \begin{minipage}{2cm}\begin{align}-\frac{G M_\rmP}{r}\nonumber \end{align}\end{minipage}     & \begin{minipage}{2cm}\begin{align}\frac{1}{s^2}\nonumber\end{align}\end{minipage}          
\\ \\ \\ \\
Plummer sphere      & \begin{minipage}{2cm}\begin{align}-\frac{G M_\rmP}{\sqrt{r^2+r^2_\rmP}}\nonumber\end{align}\end{minipage}  & \begin{minipage}{2cm}\begin{align}\frac{1}{s^2+r^2_\rmP}\nonumber\end{align}\end{minipage}         \\ \\ \\ \\
Hernquist sphere    & \begin{minipage}{2cm}\begin{align}-\frac{G M_\rmP}{r+r_\rmP}\nonumber\end{align}\end{minipage}  & 
\begin{minipage}{2cm}\begin{align}\frac{1}{r^2_\rmP-s^2}\left[\frac{r_\rmP}{\sqrt{r^2_\rmP-s^2}}\ln{\left(\frac{r_\rmP+\sqrt{r^2_\rmP-s^2}}{s}\right)}-1\right],\,\, s<r_\rmP\nonumber\end{align}\end{minipage} \\ \\ \\
& & \begin{minipage}{2cm}\begin{align}\frac{1}{s^2-r^2_\rmP}\left[1-\frac{2r_\rmP}{\sqrt{s^2-r^2_\rmP}}\tan^{-1}{\sqrt{\frac{s-r_\rmP}{s+r_\rmP}}}\right], \;\;\;\;\;\;\;\;\;\;s\geq r_\rmP\nonumber\end{align}\end{minipage}
 \\ \\ \\ \\
 NFW profile & \begin{minipage}{2cm}\begin{align}-\frac{G M_\rmP}{r}\ln{\left(1+\frac{r}{r_\rmP}\right)}\nonumber\end{align}\end{minipage} & \begin{minipage}{2cm}\begin{align}\frac{1}{s^2}\left[\ln{\left(\frac{s}{2r_\rmP}\right)}+\frac{r_\rmP}{\sqrt{r^2_\rmP-s^2}}\ln{\left(\frac{r_\rmP+\sqrt{r^2_\rmP-s^2}}{s}\right)}\right],\,\, s<r_\rmP\nonumber\end{align}\end{minipage} \\ \\ \\
& & \begin{minipage}{2cm}\begin{align}\frac{1}{s^2}\left[\ln{\left(\frac{s}{2r_\rmP}\right)}+\frac{2r_\rmP}{\sqrt{s^2-r^2_\rmP}}\tan^{-1}{\sqrt{\frac{s-r_\rmP}{s+r_\rmP}}}\right], \;\;\;\;\;\;\;\;\;\;s\geq r_\rmP\nonumber\end{align}\end{minipage}
 \\ \\ \\ \\
Isochrone potential & \begin{minipage}{2cm}\begin{align}-\frac{G M_\rmP}{r_\rmP+\sqrt{r^2+r^2_\rmP}}\nonumber\end{align}\end{minipage} & \begin{minipage}{2cm}\begin{align}\frac{1}{s^2}-\frac{r_\rmP}{s^3}\tan^{-1}\left(\frac{s}{r_\rmP}\right)\nonumber\end{align}\end{minipage} 
\\ \\ \\ \\
Gaussian potential    &  \begin{minipage}{2cm}\begin{align}-\frac{GM_\rmP}{r_\rmP}\exp{\left[-\frac{r^2}{2r^2_\rmP}\right]}\nonumber\end{align}\end{minipage}     & \begin{minipage}{2cm}\begin{align}\frac{\sqrt{\pi}}{r^2_\rmP}\exp{\left[-\frac{s^2}{2r^2_\rmP}\right]}\nonumber\end{align}\end{minipage}         \\
 \hline
\end{tabular}
   \caption{The $I(s)$ integral (see Eq.~\ref{I}) for different perturber profiles, where $s^2=x^2+{\left(b-y\right)}^2$ and $r^2=s^2+{\left(z-\vp t\right)}^2$. $M_\rmP$ and $r_\rmP$ are the mass and the scale radius of the perturber respectively. In case of the NFW profile, $M_\rmP=M_{\rm vir}/f(c)$ where $M_{\rm vir}$ is the virial mass and $f(c)=\ln{\left(1+c\right)}-c/(1+c)$, with $c=R_{\rm vir}/r_\rmP$ the concentration and $R_{\rm vir}$ the virial radius of the NFW perturber.}
\label{tab:Is}
\end{table*}

\subsection{Energy dissipation}
\label{sec:energy_straight_orbit}

An impulsive encounter imparts each subject star with an impulse $\Delta \bv(\br)$. During the encounter, it is assumed that the subject stars remain stagnant, such that their potential energy doesn't change. Hence, the energy change of each star is purely kinetic, and the total change in energy of the subject due to the encounter is given by
\begin{align}
\Delta E = \int \rmd^3 \br\, \rho_\rmS(\br) \, \Delta \varepsilon(\br) = \frac{1}{2}\int \rmd^3 \br\, \rho_\rmS(r) \, {\left(\Delta \bv\right)}^2.
\label{deltaE0}
\end{align}
Here we have assumed that the unperturbed subject is spherically symmetric, such that its density distribution depends only on $r =\left|\br\right|$, and $\Delta \varepsilon$ is given by equation~(\ref{dEstar}). We have assumed that the $\bv \cdot \Delta \bv$-term (see equation~[\ref{dEstar}]) in $\Delta \varepsilon$ vanishes, which is valid for any static, non-rotating, spherically symmetric subject. Plugging in the expression for $\Delta \bv$ from equation~(\ref{deltav1}), and substituting $x=r\sin{\theta}\cos{\phi}$ and $y=r\sin{\theta}\sin{\phi}$, we obtain
\begin{align}
\Delta E &= 2{\left(\frac{GM_\rmP}{\vp}\right)}^2 \int_0^{\infty} \rmd r\, r^2 \rho_\rmS(r) \calJ(r,b)\,,
\label{deltaE1}
\end{align}
where
\begin{align}
&\calJ(r,b)=\int_0^{\pi} \rmd \theta \sin{\theta} \int_0^{2\pi} \rmd\phi\,s^2 I^2(s)\,,
\label{J}
\end{align}
with $s^2 = x^2 + {\left(b-y\right)}^2 = r^2\sin^2{\theta} + b^2 - 2\, b\, r \sin{\theta}\sin{\phi}$. 

The above expression of $\Delta E$ includes the kinetic energy gained by the COM of the galaxy. From equation~(\ref{deltav1}), we find that the COM gains a velocity 

\begin{align}
\Delta \bv_{\rm CM} &= \frac{1}{M_\rmS}\, \int_0^\infty \rmd r\,r^2\rho_\rmS(r)\int_0^{\pi}\rmd\theta \sin{\theta}\int_0^{2\pi}\rmd \phi\, \Delta \bv = \frac{2GM_\rmP}{\vp M_\rmS}\int_0^\infty \rmd r\,r^2\rho_\rmS(r)\calJ_{\rm CM}(r,b)\,\hat{\by}\,,
\label{deltavCM}
\end{align}
where $\calJ_{\rm CM}(r,b)$ is given by
\begin{align}
&\calJ_{\rm CM}(r,b)=\int_0^{\pi}\rmd\theta \sin{\theta} \int_0^{2\pi}\rmd \phi\, I(s)\left[b-r\sin{\theta}\sin{\phi}\right]\,.
\label{J_CM}
\end{align}
Note that $\Delta \bv_{\rm CM}$ is not the same as the velocity impulse (equation~[\ref{deltav1}]) evaluated at $\br = (0,0,0)$ since we consider perturbations up to all orders. From $\Delta \bv_{\rm CM}$, the kinetic energy gained by the COM can be obtained as follows
\begin{align}
\Delta E_{\rm CM} =\frac{1}{2}M_\rmS {\left(\Delta v_{\rm CM}\right)}^2=2{\left(\frac{GM_\rmP}{\vp}\right)}^2\calV(b),
\label{deltaECM}
\end{align}
where
\begin{align}
\calV(b)=\frac{1}{M_\rmS}{\left[\int_0^{\infty}\rmd r\, r^2 \rho_\rmS(r)\calJ_{\rm CM}(r,b)\right]}^2.
\label{V}
\end{align}

We are interested in obtaining the gain in the {\it internal} energy of the galaxy. Therefore we have to subtract the energy gained by the COM from the total energy gained, which yields the following expression for the internal energy change
\begin{align}
\Delta E_{\rm int} &= \Delta E - \Delta E_{\rm CM}=2{\left(\frac{GM_\rmP}{\vp}\right)}^2 \left[\int_0^{\infty}\rmd r\,r^2\rho_\rmS(r)\calJ(r,b)-\calV(b)\right]\,.
\label{delEint1}
\end{align}

As we show in Appendix~\ref{app:asymptote}, equation~(\ref{delEint1}) has the correct asymptotic behaviour in both the large $b$ and small $b$ limits. For large $b$ it reduces to an expression that is similar to, but also intriguingly different from the standard expression obtained using the DTA, while for $b=0$ it reduces to the expression for a head-on encounter (case C in  Table~\ref{tab:comparison}).

\section{Special cases}
\label{sec:special}

In this section we discuss two special cases of perturbers for which the expression for the impulse is analytical, and for which the expression for the internal energy change of the subject can be significantly simplified.

\subsection{Plummer perturber}
\label{sec:plummer_straight_orbit}

The first special case to be considered is that of a \cite{Plummer.11} sphere perturber, the potential and $I(s)$ of which are given in Table~\ref{tab:Is}. Substituting the latter in equation~(\ref{J}) and analytically computing the $\phi$ integral yields
\begin{align}
\calJ(r,b) &= \int_0^{\pi} \rmd \theta \sin{\theta} \int_0^{2\pi} \rmd\phi\,\frac{s^2}{{\left(s^2+r^2_\rmP\right)}^2} =
4\pi\int_0^{1} \rmd \psi\,\frac{{\left(r^2-b^2-r^2\psi^2\right)}^2+r^2_\rmP\left(r^2+b^2-r^2\psi^2\right)}{{\left[{\left(r^2-b^2+r^2_\rmP-r^2\psi^2\right)}^2+4r^2_\rmP b^2\right]}^{3/2}}\,,
\label{J_plummer}
\end{align}
where $s^2 = r^2\sin^2{\theta} +b^2 - 2\,b\,r\sin{\theta}\sin{\phi}$ and $\psi=\cos{\theta}$. Similarly substituting the expression for $I(s)$ in equation~(\ref{J_CM}) yields
\begin{align}
&\calJ_{\rm CM}(r,b) = \frac{2\pi}{b} \int_0^1 \rmd \psi\,\left[1-\frac{r^2-b^2+r^2_\rmP-r^2\psi^2}{\sqrt{{\left(r^2-b^2+r^2_\rmP-r^2\psi^2\right)}^2+4r^2_\rmP b^2}}\right],
\label{J_CM_plummer}
\end{align}
which can be substituted in equation~(\ref{V}) to obtain $\calV(b)$. Both these expressions for $\calJ(r,b)$ and $\calJ_{\rm CM}(r,b)$ are easily evaluated using straightforward quadrature techniques. Finally, upon substituting $\calJ$ and $\calV$ in equation~(\ref{delEint1}), we obtain the internal energy change $\Delta E_{\rm int}$ of the subject.
\begin{figure*}
\includegraphics[width=1\textwidth]{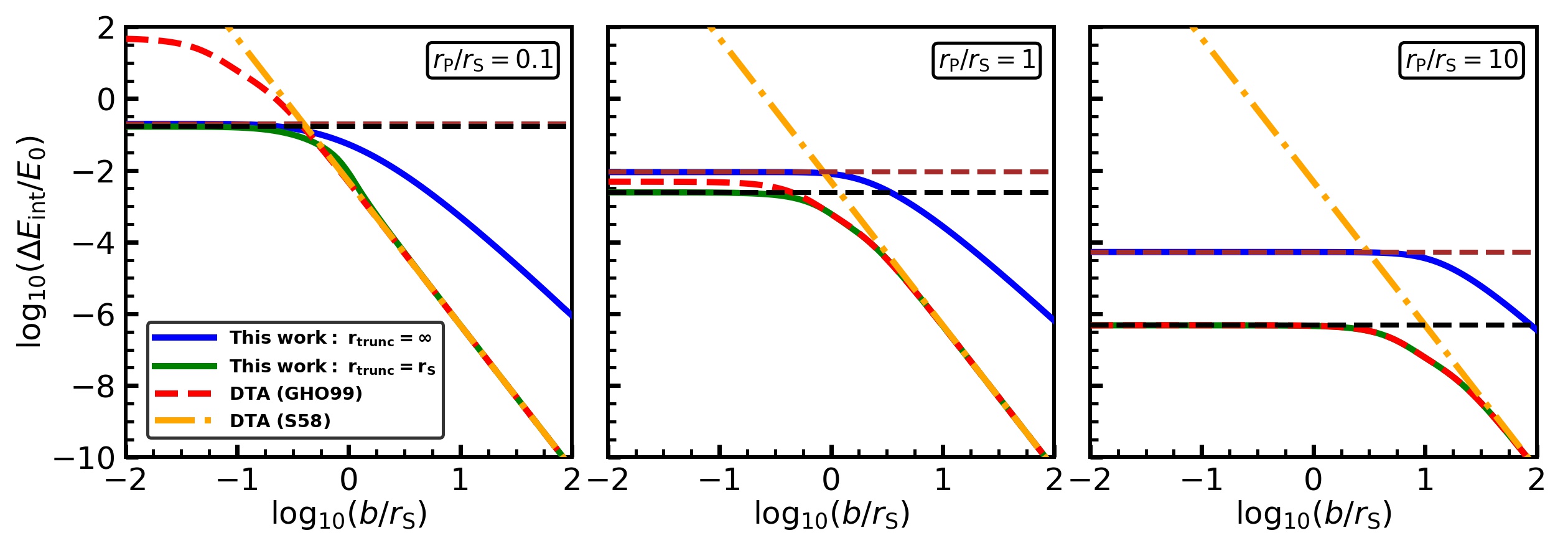}
   \caption{Impulsive heating for encounters along straight-line orbits: Each panel plots $\Delta E_{\rm int}$ in units of $E_0 = 8 \pi \, (G M_\rmP / \vp)^2 \, (M_\rmS/r^2_\rmS)$ as a function of the impact parameter $b$ in units of $r_\rmS$. Perturber and subject are modelled as Plummer and Hernquist spheres, respectively, with different panels showing results for different ratios of their characteristic radii, as indicated. The solid blue and green lines indicate $\Delta E_{\rm int}$ for infinitely extended and truncated ($r_{\rm trunc}=r_\rmS$) subjects, respectively, computed using our generalized framework (equation[~\ref{delEint1}]). The red, dashed and the orange, dot-dashed lines indicate the $\Delta E_{\rm int}$ for the truncated subject obtained using the DTA of GHO99 and S58, respectively. The brown and black dashed horizontal lines mark the head-on encounter limits for the infinite and the truncated subjects, respectively. Note that the asymptotic fall-off for the infinitely extended case (solid blue) is shallower than for the truncated case (solid green), which approaches the distant tide limit (dashed red and dot-dashed orange) for large $b$ and saturates to the head-on encounter limit for small $b$. Also note that the GHO99 approximation is in good agreement with the general result as long as the DTA is valid (i.e., $b/r_\rmS$ is large), and/or $r_\rmP$ is significantly larger than $r_\rmS$.}
\label{fig:straight_orbit}
\end{figure*}
Fig.~\ref{fig:straight_orbit} plots the resulting $\Delta E_{\rm int}$, in units of $8\pi{\left(G M_\rmP/\vp\right)}^2\left(M_\rmS/r^2_\rmS\right)$, as a function of the impact parameter, $b$, for a spherical subject with a \cite{Hernquist.90} density profile. Different panels correspond to different ratios of the characteristic radii of the perturber, $r_\rmP$, and the subject, $r_\rmS$, as indicated. Solid blue lines  indicate the $\Delta E_{\rm int}$ obtained using our non-perturbative method (equation~[\ref{delEint1}]) for an infinitely extended subject, while the solid green lines show the corresponding results for a subject truncated at $r_\rmS$. For comparison, the red, dashed and orange, dot-dashed lines show the  $\Delta E_{\rm int}$ obtained using the DTA of S58 and GHO99 (cases A and B in Table~\ref{tab:comparison}), respectively, also assuming a Hernquist subject truncated at $r_\rmS$. Finally, the black and brown horizontal, dashed lines mark the values of $\Delta E_{\rm int}$ for a head-on encounter obtained using the expression of \citet{vdBosch.etal.18a} (case C in  Table~\ref{tab:comparison}) for a truncated and infinitely extended subject, respectively.

Note that $\Delta E_{\rm int}$ for the infinitely extended subject has a different asymptotic behaviour for large $b$ than the truncated case. In fact $\Delta E_{\rm int} \propto b^{-3}$ in the case of an infinitely extended Hernquist subject (when using our non-perturbative formalism), whereas $\Delta E_{\rm int} \propto b^{-4}$ for a truncated subject (see \S\ref{sec:asymptote_tidal} for more details).

For large impact parameters, our non-perturbative $\Delta E_{\rm int}$ for the truncated case (solid green line) is in excellent agreement with the DTA of S58 and GHO99, for all three values of $r_\rmP/r_\rmS$. In the limit of small $b$, though, the different treatments yield very different predictions; whereas the $\Delta E_{\rm int}$ computed using the method of S58 diverges as $b^{-4}$, the correction of GHO99 causes $\Delta E_{\rm int}$ to asymptote to a finite value as $b \rightarrow 0$, but one that is significantly larger than what is predicted for a head-on encounter (at least when $r_\rmP < r_\rmS$). We emphasize, though, that both the S58 and GHO99 formalisms are based on the DTA, and therefore not valid in this limit of small $b$. In contrast, our non-perturbative method is valid for all $b$, and nicely asymptotes to the value of a head-on encounter in the limit $b \rightarrow 0$.

It is worth pointing out that the GHO99 formalism yields results that are in excellent agreement with our fully general, non-perturbative approach when $r_\rmP/r_\rmS \gg 1$, despite the fact that it is based on the DTA. However, this is only the case when the subject is truncated at a sufficiently small radius $r_{\rm trunc}$. Recall that the DTA yields that $\Delta E_{\rm int} \propto \langle r^2 \rangle$ (see Table~\ref{tab:comparison}), which diverges unless the subject is truncated or the outer density profile of the subject has $\rmd\log\rho_\rmS/\rmd\log r < -5$. In contrast, our generalized formalism yields a finite $\Delta E_{\rm int}$, independent of the density profile of the subject.

\begin{figure}
\includegraphics[width=0.9\textwidth]{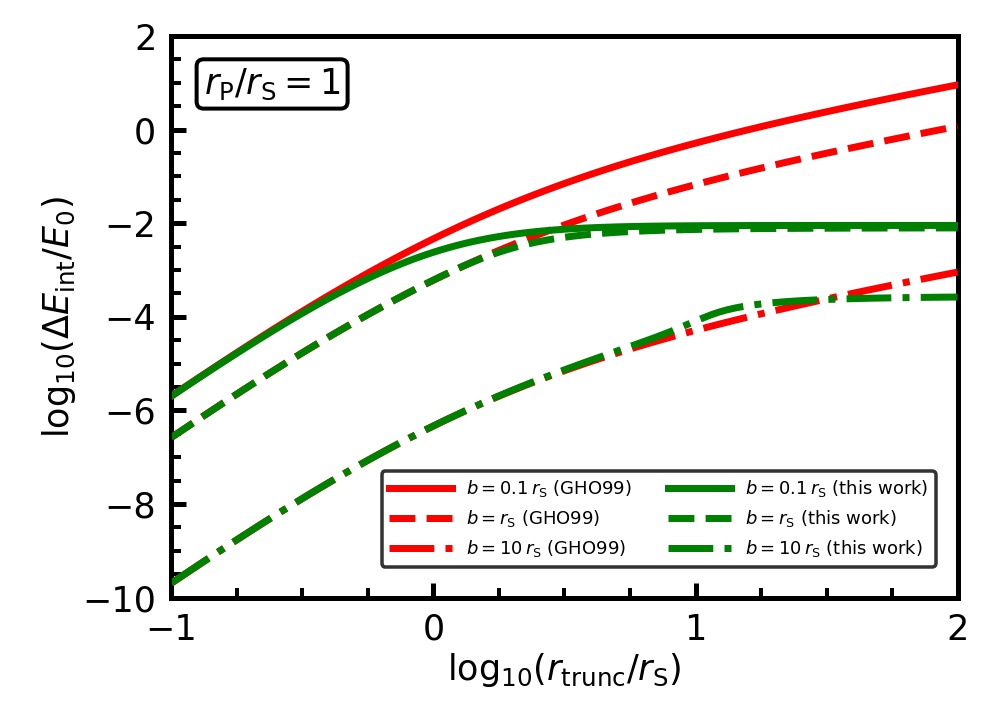}
   \caption{The increase in internal energy, $\Delta E_{\rm int}$, in units of $E_0=8\pi{\left(G M_\rmP/\vp\right)}^2\left(M_\rmS/r^2_\rmS\right)$, of a truncated Hernquist sphere due to an impulsive encounter with a Plummer sphere perturber with $r_\rmP/r_\rmS=1$ along a straight-line orbit. Results are shown as a function of the subject's truncation radius, $r_{\rm trunc}$, in units of $r_\rmS$, for three values of the impact parameter, $b/r_\rmS$, as indicated. Green and red lines correspond to the  $\Delta E_{\rm int}$ computed using our generalized framework and the DTA of GHO99, respectively.}
\label{fig:straight_orbit_rc}
\end{figure}

This is illustrated in Fig.~\ref{fig:straight_orbit_rc} which plots $\Delta E_{\rm int}$, again in units of $8\pi{\left(G M_\rmP/\vp\right)}^2\left(M_\rmS/r^2_\rmS\right)$, as a function of $r_{\rm trunc}/r_\rmS$ for a Plummer perturber and a truncated Hernquist subject with $r_\rmP/r_\rmS=1$. Results are shown for three different impact parameters, as indicated. The green and red lines indicate the $\Delta E_{\rm int}$ obtained using our general formalism and that of GHO99, respectively. Note that the results of GHO99 are only in good agreement with our general formalism when the truncation radius is small and/or the impact parameter is large. 
\subsection{Point mass perturber}
\label{sec:point_mass_straight_orbit}

The next special case to discuss is that of a point mass perturber, which one can simply obtain by taking the results for a spherical Plummer perturber discussed above in the limit $r_\rmP \rightarrow 0$. In this limit the $\calJ$ integral of equation~(\ref{J_plummer}) can be computed analytically and substituted in equation~(\ref{deltaE1}) to yield 
\begin{align}
&\Delta E = 4\pi{\left(\frac{GM_\rmP}{\vp}\right)}^2 \int_0^{\infty} \rmd r\, r^2 \rho_\rmS(r) \int_0^{\pi} \rmd \theta \frac{\sin{\theta}}{\left|b^2-r^2\sin^2{\theta}\right|}\,.
\label{deltaE1_point}
\end{align}
The same applies to the $\calJ_{\rm CM}$ integral of equation~(\ref{J_CM_plummer}), which yields the following COM velocity
\begin{align}
\Delta \bv_{\rm CM} = \frac{2GM_\rmP}{\vp M_\rmS}\frac{M_{\rm enc}(b)}{b}\hat{\by}\,,
\label{deltavCM_ptmass}
\end{align}
where $M_{\rm enc}(b)$ is the galaxy mass enclosed within a cylinder of radius $b$, and is given by
\begin{align}
&M_{\rm enc}(b) = 4\pi \left[\int_0^b \rmd r\, r^2\rho_\rmS(r) + \int_b^\infty \rmd r\,r^2\rho_\rmS(r)\left(1-\sqrt{1-\frac{b^2}{r^2}}\right) \right].
\label{Menc}
\end{align}
Therefore, the kinetic energy gained by the COM in the encounter can be written as
\begin{align}
\Delta E_{\rm CM} = \frac{1}{2 M_\rmS} {\left[\frac{2GM_\rmP}{\vp}\frac{M_{\rm enc}(b)}{b}\right]}^2.
\label{deltaECM_ptmass}
\end{align}
Subtracting this from the expression for $\Delta E$ given in equation~(\ref{deltaE1_point}) and analytically computing the $\theta$ integral yields the following expression for the internal energy change
\begin{align}
\Delta E_{\rm int} &= 8\pi {\left(\frac{GM_\rmP}{\vp}\right)}^2 \int_0^{r_{\rm trunc}} \rmd r\,\rho_\rmS(r)\left[\frac{r}{\sqrt{b^2-r^2}}\tan^{-1}{\left(\frac{r}{\sqrt{b^2-r^2}}\right)}-\frac{r^2}{b^2}\right]\,.
\label{deltaEint_ptmass}
\end{align}
Here we assume the subject to be truncated at some $r_{\rm trunc}<b$, and therefore $M_{\rm enc}(b)=M_\rmS$. If $r_{\rm trunc}>b$, then the point perturber passes through the subject and imparts an infinite impulse in its neighbourhood, which ultimately leads to a divergence of $\Delta E_{\rm int}$. 

Note that the term in square brackets tends to $\frac{2}{3} (r/b)^4$ in the limit $r \ll b$. Hence, the above expression for $\Delta E_{\rm int}$ reduces to the standard distant tide expression of S58, given in equation~(\ref{dESpitzer}), as long as $b \gg r_{\rm trunc}$.
Unlike S58 though, the above expression for $\Delta E_{\rm int}$ is applicable for any $b>r_{\rm trunc}$, and is therefore a generalization of the former.

\subsection{Other perturbers}
\label{sec:other_perturbers}

The Plummer and point-mass perturbers discussed above are somewhat special in that the corresponding expression for the impulse is sufficiently straightforward that the expression for $\Delta E_{\rm int}$ (equation~[\ref{delEint1}]) simplifies considerably. For the other perturber profiles listed in Table~\ref{tab:Is}, $\Delta E_{\rm int}$ is to be computed by numerically evaluating the $\calJ$ and $\calJ_{\rm CM}$ integrals given in equations~(\ref{J}) and~(\ref{J_CM}), respectively. We provide a Python code, {\tt NP-impulse}\footnote{\url{https://github.com/uddipanb/NP-impulse}}, that does so, and that can be used to compute  $\Delta E_{\rm int}(b,v)$ for a variety of (spherical) perturber and subject  profiles. We emphasize that the results are in good agreement with the estimates of GHO99, which are based on the DTA, when (i) the perturber is sufficiently extended (i.e., $r_\rmP > r_\rmS$), and (ii) the subject is truncated at a radius $r_{\rm trunc} < b$. When these conditions are not met, the GHO99 formalism typically significantly overpredicts $\Delta E_{\rm int}$ at small impact parameters. Our more general formalism, on the other hand, remains valid for any $b$ and any $r_{\rm trunc}$ (including no truncation), and smoothly asymptotes to the analytical results for a head-on encounter. 

\section{Encounters along eccentric orbits}
\label{sec:eccentric_orbit}

In the previous sections we have discussed how a subject responds to a perturber that is moving along a straight-line orbit. The assumption of a straight-line orbit is only reasonable in the highly impulsive regime, when $\vp \gg \sigma$. Such situations do occur in astrophysics (i.e., two galaxies having an encounter within a cluster, or a close encounter between two globular clusters in the Milky Way). However, one also encounters cases where the encounter velocity is largely due to the subject and perturber accelerating each other (i.e., the future encounter of the Milky Way and M31), or in which the subject is orbiting within the potential of the perturber (i.e., M32 orbiting M31). In these cases, the assumption of a straight-line orbit is too simplistic.  In this section we therefore generalize the straight-line orbit formalism developed in \S\ref{sec:straight_orbit}, to the case of subjects moving on eccentric orbits within the perturber potential. Our approach is similar to that in GHO99, except that we refrain from using the DTA, i.e., we do not expand the perturber potential in multi-poles and we do not assume that  $r_\rmP \gg r_\rmS$. Rather our formalism is applicable to any sizes of the subject and the perturber. In addition, our formalism is valid for any impact parameter (which here corresponds to the pericentric distance of the eccentric orbit), whereas the formalism of GHO99 is formally only valid for $b \gg r_\rmS$. However, like GHO99, our formalism is also based on the impulse approximation, which is only valid as long as the orbit is sufficiently eccentric such that the encounter time, which is of order the timescale of pericentric passage, is shorter than the average orbital timescale of the subject stars. Since the stars towards the central part of the subject orbit much faster than those in the outskirts, the impulse approximation can break down for stars near the centre of the subject, for whom the encounter is adiabatic rather than impulsive. As discussed in \S\ref{sec:adiabatic_shielding}, we can take this `adiabatic shielding' into account using the adiabatic correction formalism introduced by \citet{Gnedin.Ostriker.99}. This correction becomes more significant for less eccentric orbits.

\subsection{Orbit characterization}
\label{sec:eccentric_orbit_character}

We assume that the perturber is much more massive than the subject ($M_\rmP \gg M_\rmS$) and therefore governs the motion of the subject. We also assume that the perturber is spherically symmetric, which implies that the orbital energy and angular momentum of the subject are conserved and that its orbit is restricted to a plane. This orbital energy and angular momentum (per unit mass) are given by
\begin{align}
E &= \frac{1}{2}\dot{R}^2+\Phi_\rmP(R)+\frac{L^2}{2R^2}, \nonumber \\
L &= R^2\dot{\theta}_\rmP,
\label{EL}
\end{align}
where $\bR$ is the position vector of the COM of the perturber with respect to that of the subject, $R= \left|\bR\right|$, and $\theta_\rmP$ is the angle on the orbital plane defined such that $\theta_\rmP = 0$ when $R$ is equal to the pericentric distance, $R_{\rm peri}$. The dots denote derivatives with respect to time. The above equations can be rearranged and integrated to obtain the following forms for $\theta_\rmP$ and $t$ as functions of $R$
\begin{align}
&\theta_\rmP(R) = \int_{1/\alpha}^{R/r_\rmP} \frac{\rmd R'}{R'^2\sqrt{2\left[\calE-\Phi'_\rmP(R')\right]/\ell^2-1/R'^2}},\nonumber \\
&t(R) = \int_{1/\alpha}^{R/r_\rmP} \frac{\rmd R'}{\ell\sqrt{2\left[\calE-\Phi'_\rmP(R')\right]/\ell^2-1/R'^2}}.
\end{align}
Here $\alpha=r_\rmP/R_{\rm peri}$, $t$ is in units of ${\left(r^3_\rmP/GM_\rmP\right)}^{1/2}$, and $\calE=E \left(r_\rmP /GM_\rmP\right)$, $\Phi'_\rmP=\Phi_\rmP \left(r_\rmP/GM_\rmP\right)$ and $\ell=L/{\left(GM_\rmP r_\rmP\right)}^{1/2}$ are dimensionless expressions for the orbital energy, perturber potential and orbital angular momentum, respectively. The resulting orbit is a rosette, with $R$ confined between a pericentric distance, $R_{\rm peri}$, and an apocentric distance, $R_{\rm apo}$. The angle between a pericenter and the subsequent apocenter is $\theta_{\rm max}$, which ranges from $\pi/2$ for the harmonic potential to $\pi$ for the Kepler potential \citep[e.g.,][]{Binney.Tremaine.87}. The orbit's eccentricity is defined as 
\begin{align}
e = \frac{R_{\rm apo}-R_{\rm peri}}{R_{\rm apo}+R_{\rm peri}},
\end{align}
which ranges from $0$ for a circular orbit to $1$ for a purely radial orbit. Here we follow GHO99 and characterize an orbit by $e$ and $\alpha = r_\rmP/R_{\rm peri}$.

\subsection{Velocity perturbation and energy dissipation}
\label{sec:velocity_energy_eccentric_orbit}

The position vector of the perturber with respect to the subject is given by $\bR=R \cos{\theta_\rmP}\hat{\by}+R\sin{\theta_\rmP}\hat{\bz}$, where we take the orbital plane to be spanned by the $\hat{\by}$ and $\hat{\bz}$ axes, with $\hat{\by}$ directed towards the pericenter. The function $R(\theta_\rmP)$ specifies the orbit of the subject in the perturber potential and is therefore a function of the orbital parameters $\alpha$ and $e$. In the same spirit as in equation~(\ref{force}), we write the acceleration due to the perturber on a subject star located at $(x,y,z)$ from its COM as
\begin{align}
&\ba_\rmP = -\nabla \Phi_\rmP = \frac{1}{R_\rmP}\frac{\rmd\Phi_\rmP}{\rmd R_\rmP} \left[-x\hat{\bx}+\left(R\cos{\theta_\rmP}-y\right)\hat{\by}+\left(R\sin{\theta_\rmP}-z\right)\hat{\bz}\right], 
\label{force_eccentric}
\end{align}
where $R_\rmP=\sqrt{x^2+{\left(R\cos{\theta_\rmP}-y\right)}^2+{\left(R\sin{\theta_\rmP}-z\right)}^2}$ is the distance of the star from the perturber. We are interested in the response of the subject during the encounter, i.e., as the perturber moves (in the reference frame of the subject) from one apocenter to another, or equivalently from $(R_{\rm apo},-\theta_{\rm max})$ to $(R_{\rm apo},\theta_{\rm max})$. During this period, $T$, the star particle undergoes a velocity perturbation $\Delta \bv$, given by
\begin{align}
&\Delta \bv = \int_{-T/2}^{T/2} \rmd t\, \ba_\rmP = \frac{1}{L} \int_{-\theta_{\rm max}}^{\theta_{\rm max}} \rmd \theta_\rmP R^2(\theta_\rmP) \,\frac{1}{R_\rmP}\frac{\rmd\Phi_\rmP}{\rmd R_\rmP}\left[-x\hat{\bx}+\left(R\cos{\theta_\rmP}-y\right)\hat{\by}+\left(R\sin{\theta_\rmP}-z\right)\hat{\bz}\right],
\label{deltav0_eccentric}
\end{align}
where we have substituted $\theta_\rmP$ for $t$ by using the fact that $\dot{\theta}_\rmP = L/R^2$. Also, using that $L=\ell\sqrt{GM_\rmP r_\rmP}$ and $\Tilde{\Phi}_\rmP = \Phi_\rmP/(GM_\rmP)$, the above expression for $\Delta \bv$ can be more concisely written as
\begin{align}
\Delta \bv &= \sqrt{\frac{GM_\rmP}{r_\rmP}}\frac{1}{\ell(\alpha,e)}\left[-x I_1 \hat{\bx}+\left(I_2-y I_1\right) \hat{\by}+\left(I_3-z I_1\right) \hat{\bz}\right],
\label{deltav0_eccentric_conc}
\end{align}
where
\begin{align}
I_1(\br) &= \int_{-\theta_{\rm max}}^{\theta_{\rm max}} \rmd \theta_\rmP\, R^2(\theta_\rmP) \frac{1}{R_\rmP}\frac{\rmd\Tilde{\Phi}_\rmP}{\rmd R_\rmP}, \nonumber \\
I_2(\br) &= \int_{-\theta_{\rm max}}^{\theta_{\rm max}} \rmd \theta_\rmP\cos{\theta_\rmP}\, R^3(\theta_\rmP) \frac{1}{R_\rmP}\frac{\rmd\Tilde{\Phi}_\rmP}{\rmd R_\rmP}, \nonumber \\
I_3(\br) &= \int_{-\theta_{\rm max}}^{\theta_{\rm max}} \rmd \theta_\rmP\sin{\theta_\rmP}\, R^3(\theta_\rmP) \frac{1}{R_\rmP}\frac{\rmd\Tilde{\Phi}_\rmP}{\rmd R_\rmP}.
\end{align}
Note that $I_1$ has units of inverse length, while $I_2$ and $I_3$ are unitless.

Over the duration of the encounter, the COM of the subject (in the reference frame of the perturber) undergoes a velocity change 
\begin{align}
\Delta \bv_{\rm CM} & = 2 \, R_{\rm apo}\, \dot{\theta}_\rmP\vert_{\rm apo} \, \sin{\theta_{\rm max}} \, \hat{\by} = 2 \, \sqrt{\frac{GM_\rmP}{r_\rmP}} \, \alpha \, \ell(\alpha,e) \, \frac{1-e}{1+e} \, \sin{\theta_{\rm max}} \, \hat{\by}.
\end{align}
Subtracting this $\Delta \bv_{\rm CM}$ from $\Delta \bv$, we obtain the velocity perturbation $\Delta \bv_{\rm rel} = \Delta \bv - \Delta \bv_{\rm CM}$ relative to the COM of the subject, which implies a change in internal energy given by
\begin{align}
\Delta E_{\rm int} = \frac{1}{2}\int_0^\infty \rmd r\,r^2\rho_\rmS(r)\int_0^{\pi}\rmd\theta \sin{\theta}\int_0^{2\pi}\rmd \phi\, \Delta v^2_{\rm rel}.
\end{align}
Substituting the expression for $\Delta \bv$ given by equation~(\ref{deltav0_eccentric_conc}), we have that
\begin{align}
\Delta E_{\rm int} &= \frac{GM_\rmP}{2r_\rmP}\int_0^\infty \rmd r\,r^2\rho_\rmS(r)\int_0^{\pi}\rmd\theta \sin{\theta}\int_0^{2\pi}\rmd \phi\,\, \calK (\br).
\label{deltaEint_eccentric}
\end{align}
Here the function $\calK (\br)$ is given by
\begin{align}
\calK(\br) = \frac{r^2 \,I^2_1 + I'^2_2 + I^2_3 - 2\, r \,I_1  \left(I'_2\sin{\theta}\sin{\phi} + I_3\cos{\theta} \right)}{\ell^2(\alpha,e)},
\label{calK}
\end{align}
where $I'_2 = I_2 -\Delta \Tilde{v}_{\rm CM}$, with 
\begin{align}
\Delta \Tilde{v}_{\rm CM}= 2 \alpha \, \ell^2(\alpha,e) \, \frac{1-e}{1+e} \, \sin{\theta_{\rm max}}.
\end{align}

Finally, from the conservation of energy and equation~(\ref{EL}), it is straightforward to  infer that\footnote{Analytical expressions for a few specific perturber potentials are listed in Table~1 of GHO99.} 
\begin{align}
{\ell}^2(\alpha,e) = \frac{{\left(1+e\right)}^2}{2e} \, \frac{r_\rmP}{\alpha^2} \, \left[\Tilde{\Phi}_\rmP \left(\frac{r_\rmP}{\alpha} \frac{1+e}{1-e}\right) - \Tilde{\Phi}_\rmP \left(\frac{r_\rmP}{\alpha}\right) \right].
\end{align}
\begin{figure*}
\includegraphics[width=1\textwidth]{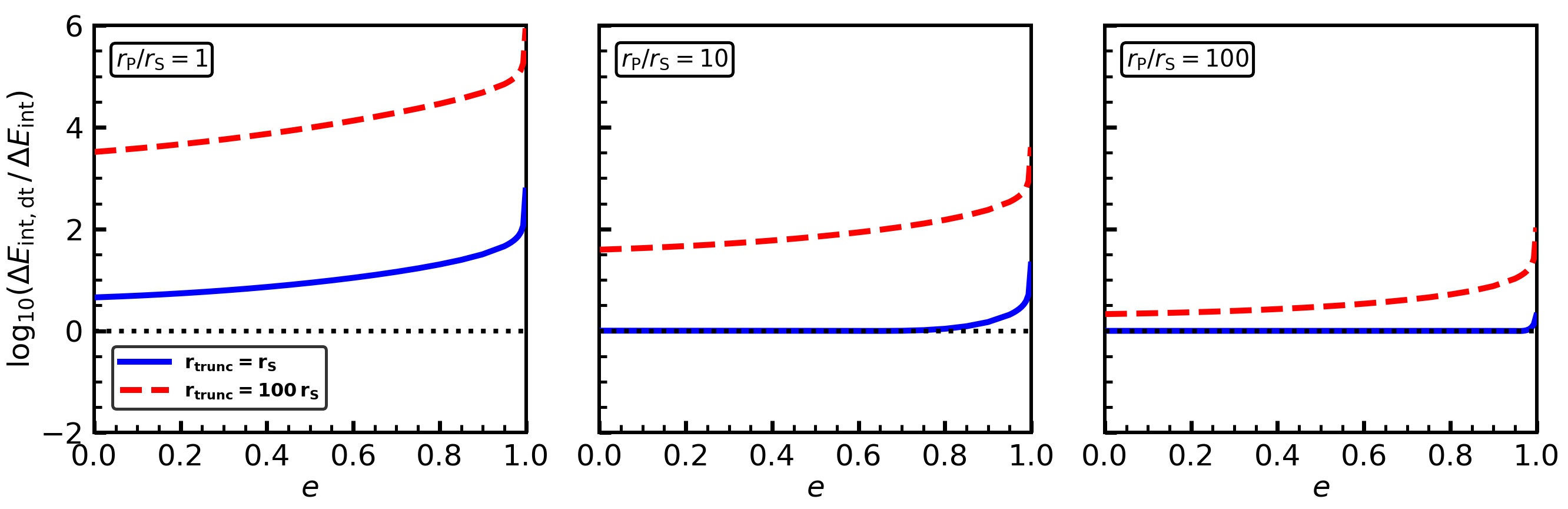}
   \caption{Impulsive heating for encounters along eccentric orbits:  Blue, solid and red, dashed lines indicate the ratio of $\Delta E_{\rm int}$ computed using the DTA of GHO99 ($\Delta E_{\rm {int,dt}}$) to that computed using our general formalism (equation~[\ref{deltaEint_eccentric_ad_corr}]) as a function of the orbital eccentricity, $e$, for cases in which the spherical Hernquist subject is truncated at $r_{\rm trunc}=r_\rmS$ and $100\,r_\rmS$, respectively. In each case, the orbital energy is $E = -0.7GM_\rmP/r_\rmP$, and the perturber is modelled as a Hernquist sphere with $M_\rmp = 1000 M_\rmS$ (here $M_\rmS$ is the subject mass enclosed within its truncation radius). Different panels correspond to different $r_\rmP/r_\rmS$, as indicated.}
\label{fig:eccentric_orbit}
\end{figure*}
\subsection{Adiabatic correction}
\label{sec:adiabatic_shielding}

The expression for the internal energy change of the subject derived in the previous section (equation~[\ref{deltaEint_eccentric}]) is based on the impulse approximation. This assumes that during the encounter the stars only respond to the perturbing force and not to the self-gravity of the subject. However, unless the encounter speed is much larger than the internal velocity dispersion of the subject, this is a poor approximation towards the center of the subject, where the dynamical time of the stars, $t_{\rm dyn}(r) \propto \left[G \rho_\rmS(r)\right]^{-1/2}$ can be comparable to, or even shorter than, the time scale of the encounter $\tau$. For such stars the encounter is not impulsive at all; in fact, if $t_{\rm dyn}(r) \ll \tau$ the stars respond to the encounter adiabatically, such that the net effect of the encounter leaves their energy and angular momentum invariant. In this section we modify the expression for $\Delta E_{\rm int}$ derived above by introducing an adiabatic correction to account for the fact that the central region of the subject may be  `adiabatically shielded' from the tidal shock.

We follow \citet{Gnedin.Ostriker.99} who, using numerical simulations and motivated by \cite{Weinberg.94a, Weinberg.94b}, find that the ratio of the actual, average energy change $\langle \Delta E \rangle(r)$ for subject stars at radius $r$ to that predicted by the impulse approximation, is given by
\begin{align}
\calA(r) = \left[ 1 + \omega^2(r) \tau^2 \right]^{-\gamma}.
\label{ad_corr}
\end{align}
Here $\tau$ is the shock duration, which is of order the timescale of pericentric passage, i.e.,
\begin{align}
\tau \sim \frac{1}{\dot{\theta}_\rmP\vert_{\rm peri}} = \sqrt{\frac{r^3_\rmP}{G M_\rmP}}\frac{1}{\alpha^2\,\ell(\alpha,e)}\,,
\end{align}
and $\omega(r) = \sigma(r)/r$ is the frequency of subject stars at radius $r$, with $\sigma(r)$ the isotropic velocity dispersion given by
\begin{align}
\sigma^2(r) = \frac{1}{\rho_\rmS(r)} \int_r^{\infty} \rmd r' \, \rho_\rmS(r') \, \frac{\rmd\Phi_\rmS}{\rmd r'}.
\end{align}

For the power-law index $\gamma$, \citet{Gnedin.Ostriker.99} find that it obeys
\begin{align}
\gamma&=
    \begin{cases}
     2.5, &\tau \lesssim t_{\rm dyn} \\
     1.5, &\tau \gtrsim 4\,t_{\rm dyn},
    \end{cases}
\end{align}
where
\begin{align}
t_{\rm dyn}=\sqrt{\frac{\pi^2 r^3_\rmh}{2GM_\rmS}}
\end{align}
is the dynamical time at the half mass radius $r_\rmh$ of the subject. In what follows we therefore adopt
\begin{align}
 \gamma = 2 - 0.5\,\rm erf \left(\frac{\tau-2.5\,t_{\rm dyn}}{0.7\,t_{\rm dyn}}\right)
\end{align}
as a smooth interpolation between the two limits. Implementing this adiabatic correction, we arrive at the following final expression for the internal energy change of the subject during its encounter with the perturber
\begin{align}
\Delta E_{\rm int} = \frac{GM_\rmP}{2r_\rmP} \int_0^\infty \rmd r \, r^2 \, \rho_\rmS(r) \, \calA(r) \, \int_0^{\pi} \rmd\theta \, \sin{\theta} \, \int_0^{2\pi} \rmd \phi\,\calK (\br)\,.
\label{deltaEint_eccentric_ad_corr}
\end{align}

We caution that the adiabatic correction formalism of \citet{Gnedin.Ostriker.99} has not been tested in the regime of small impact parameters. In addition, ongoing studies suggest that equation~(\ref{ad_corr}) may require a revision for the case of extensive tides \citep[][]{Martinez-Medina.etal.20}. Hence, until an improved and well-tested formalism for adiabatic shielding is developed, the results in this subsection have to be taken with a grain of salt. However, as long as the adiabatic correction remains a function of radius only, equation~(\ref{deltaEint_eccentric_ad_corr}) remains valid.

In Fig.~\ref{fig:eccentric_orbit}, we compare this $\Delta E_{\rm int}$ with that computed using the DTA of GHO99, which can be written in the form of equation~(\ref{deltaEint_eccentric_ad_corr}) but with $\calK(\br)$ replaced by
\begin{align}
    \calK_{\rm GHO}(\br) = \left({r \over r_\rmP} \right)^2 \, {(B_1 - B_3)^2\sin^2{\theta}\sin^2{\phi} + (B_2 - B_3)^2\cos^2{\theta} + B_3^2\sin^2{\theta}\cos^2{\phi} \over \, \ell^2(\alpha,e)}
\label{calK_GHO}
\end{align}
with $B_1$, $B_2$ and $B_3$ integrals, given by equations~(36), (37) and~(38) in GHO99, that carry information about the perturber profile and the orbit. The lines show the ratio of $\Delta E_{\rm int}$ computed using GHO99's DTA and that computed using our formalism (equations~[\ref{deltaEint_eccentric_ad_corr}] and~[\ref{calK}]) as a function of the orbital eccentricity $e$, and for an orbital energy $E=-0.7GM_\rmP/r_\rmP$.  Both the perturber and the subject are modelled as Hernquist spheres. Solid blue and dashed red lines correspond to cases in which the subject is truncated at $r_{\rm trunc} = r_\rmS$ and $100\,r_\rmS$, respectively, while different panels correspond to different ratios of $r_\rmP/r_\rmS$, as indicated.

The GHO99 results are in excellent agreement with our more general formalism when $r_{\rm trunc}=r_\rmS$ and $r_\rmP / r_\rmS \gg 1$. Note, though, that the former starts to overpredict $\Delta E_{\rm int}$ in the limit $e \to 1$. The reason is that for higher eccentricities, the pericentric distance becomes smaller and the higher-order multipoles of the perturber potential start to contribute more. Since the DTA truncates $\Phi_\rmP$ at the quadrupole, it becomes less accurate. As a consequence, the GHO99 results actually diverge in the limit $e \to 1$, while the $\Delta E_{\rm int}$ computed using our fully general formalism continues to yield finite values. The agreement between our $\Delta E_{\rm int}$ and that computed using the GHO99 formalism becomes worse for smaller $r_\rmP/r_\rmS$ and larger $r_{\rm trunc}$. When $r_\rmP/r_\rmS=1$ (left-hand panel), GHO99 overpredicts $\Delta E_{\rm int}$ by about one to two orders of magnitude when $r_{\rm trunc} = r_\rmS$, which increases to 3 to 5 orders of magnitude for  $r_{\rm trunc} = 100\,r_\rmS$. Once again, this sensitivity to $r_{\rm trunc}$ has its origin in the fact that the integral $\int_0^{r_{\rm trunc}}\rmd r\,r^4\,\rho_\rmS(r)\,\calA(r)$ diverges as $r_{\rm trunc} \to \infty$ for the Hernquist $\rho_\rmS(r)$ considered here.

\section{Mass Loss due to Tidal Shocks in Equal Mass Encounters}
\label{sec:mass_loss}

\begin{figure}
\centering
\hspace{-1mm}
\subfloat{\includegraphics[width=1\textwidth]{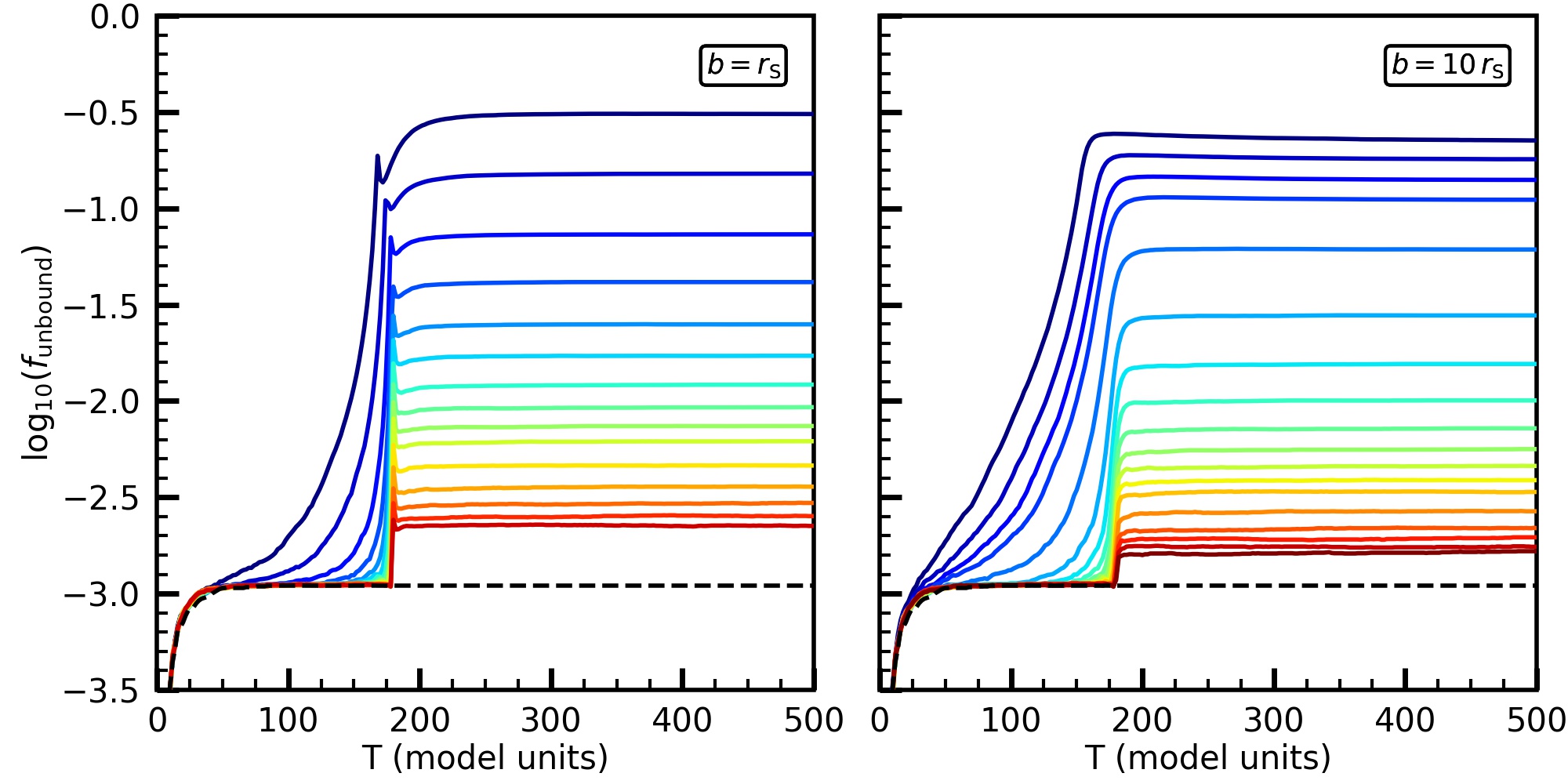}}\\
\subfloat{\includegraphics[width=1\textwidth]{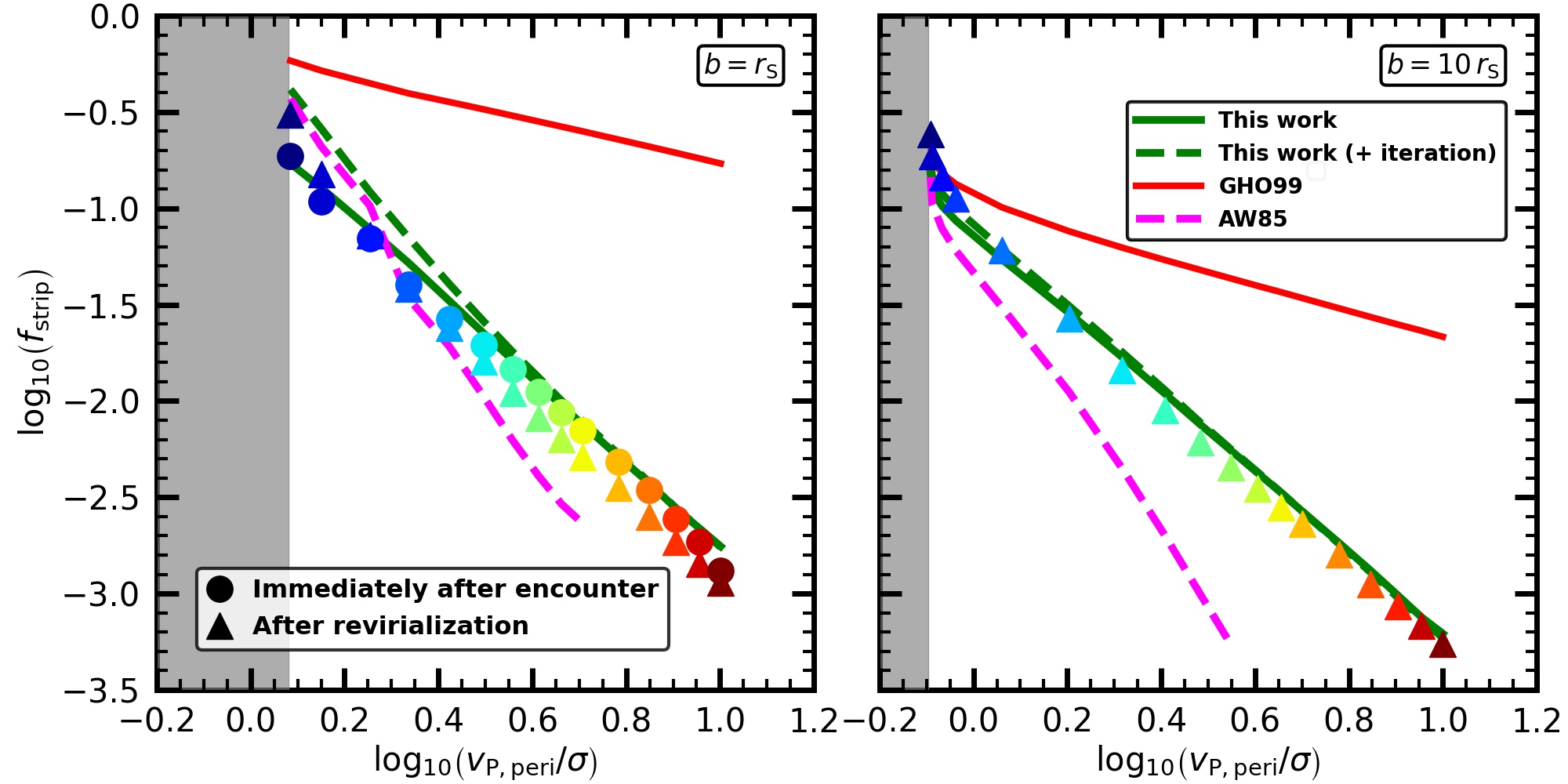}}
\caption{Comparison of numerical simulations with Monte Carlo predictions for the amount of mass loss induced by a tidal shock resulting from a penetrating encounter between two identical Hernquist spheres with impact parameter $b=r_\rmS$ (left panels) and $10\,r_\rmS$ (right panels). Upper panels show the time evolution of the unbound mass fraction, $f_{\rm unbound}$, in N-body simulations of encounters with different initial encounter velocities, $\vp$, ranging from $0.4\,\sigma$ ($0.7\,\sigma$) to $10\,\sigma$ for $b=10\,r_\rmS$ ($b=r_\rmS$), color coded from blue to red. The solid circles and triangles in the lower panels show the corresponding stripped mass fractions, $f_{\rm strip}$, as a function of $v_{\rm P,peri}/\sigma$ immediately following the encounter and after revirialization, respectively. For comparison, the solid green lines show the predictions from our general formalism for computing the impulse (equation~[\ref{deltav1}]). The solid red lines denote the predictions obtained using the DTA of GHO99. We emphasize that the DTA is not valid for penetrating encounters, and that the red lines are merely included for comparison. The green dashed lines show the predictions obtained using an iterative approach to determine the maximal subset of self-bound particles following the encounter. Finally, the dashed magenta lines show the predictions based on the fitting formula of \protect\cite{Aguilar.White.85} (AW85), which is based on a similar iterative approach, but applied to less extended objects. The grey shaded regions indicate the encounter velocities that result in tidal capture. See text for details and discussion.}
\label{fig:mass_loss}
\end{figure}

In this section we present an astrophysical application of our generalized formalism. We consider penetrating gravitational encounters between two spherical systems. In what follows we refer to them as (dark matter) haloes, though they could represent stellar systems equally well. We investigate the amount of mass loss to be expected from such encounters, and, in order to validate our formalism, compare its predictions to the results from $N$-body simulations.

\subsection{Numerical Simulations}
\label{sec:sim}

We simulate encounters between two identical, spherical \cite{Hernquist.90} haloes, whose initial density profile and potential are given by
\begin{equation}
\rho_\rmS(r) = \rho_0 \, \left({r \over r_\rmS}\right)^{-1} \, \left(1 + {r\over r_\rmS}\right)^{-3}, \;\;\;\;\;\;\;\;{\rm and}\;\;\;\;\;\;\;\;
\Phi_\rmS(r) = -\frac{G M_\rmS}{r+r_\rmS}\,,
\end{equation}
where $\rho_0=M_\rmS/2\pi r^3_\rmS$. Throughout we adopt model units in which the gravitational constant, $G$, the characteristic scale radius, $r_\rmS$, and the initial mass of each halo, $M_\rmS$, are all unity. Each halo is modelled using $N_\rmp = 10^5$ particles, whose initial phase-space coordinates are sampled from the ergodic distribution function $f = f(E)$ under the assumption that the initial haloes have isotropic velocity distributions. For practical reasons, each Hernquist sphere is truncated at $r_{\rm trunc}=1000\,r_\rmS$, which encloses 
99.8\% of $M_\rmS$. 

The haloes are initialized to approach each other with an impact parameter $b$, and an initial velocity $\vp$. We examine the cases of $b=r_\rmS$ and $10\,r_\rmS$. Initially the haloes are placed at large distance from each other, such that, depending on $\vp$, they always reach closest approach at $t \sim 200$. The simulation is continued up to $t = 500$, allowing the haloes sufficient time to re-virialize following the encounter.

The encounter is followed using the $N$-body code  {\tt treecode}, written by Joshua Barnes, which uses a \cite{Barnes.Hut.86} octree to compute accelerations based on a multipole expansion up to quadrupole order, and a second order leap-frog integration scheme to solve the equations of motion. Since we use fixed time step, our integration scheme is fully symplectic. Forces between particles are softened using a simple Plummer softening. Throughout we adopt a time step of $\Delta t = 0.02$ and a softening length of $\varepsilon_{\rm soft} = 0.05$. These values ensure that the halo in isolation remains in equilibrium for well over $10 \Gyr$. For each $b$ we have run simulations for different values of $\vp/\sigma$ in the range $[0.4, 10]$. Here $\sigma=\sqrt{GM_\rmP/r_\rmS}$ is the characteristic internal velocity dispersion of the Hernquist halo.

For each of the two haloes, we measure its fraction of unbound particles $f_{\rm unbound}$ using the iterative method described in  \citet{vdBosch.etal.18b}.\footnote{Note though, that unlike in that paper, we determine the centre of mass position and velocity of each halo using all the bound particles, rather than only the 5 percent most bound particles. We find that this yields more stable results.} In the upper left (right) panel of Fig.~(\ref{fig:mass_loss}), we plot $f_{\rm unbound}$ (averaged over the two haloes) as a function of time for $b=r_\rmS$ ($10\,r_\rmS$). The solid curves correspond to different $v_\rmP$ ranging from $0.7$ to $10\,\sigma$ for $b=r_\rmS$ and from $0.4$ to $10\,\sigma$ for $b=10\,r_\rmS$, color coded from blue to red. The black dashed curve indicates the $f_{\rm unbound}$ for an isolated halo. Each halo is subject to an initial unbinding of $\sim 0.1\%$ of its mass due to numerical round-off errors in the force computation for particles in the very outskirts. The mass loss induced by the encounter occurs in two steps. First the tidal shock generated by the encounter unbinds a subset of particles. Subsequently, the system undergoes re-virialization during which the binding energies of individual particles undergo changes. This can both unbind additional particles, but also rebind particles that were deemed unbound directly following the tidal shock. Re-virialization is a more pronounced effect for more penetrating encounters, i.e., for $b=r_\rmS$. In this case, $f_{\rm unbound}$ increases steeply during the encounter and exhibits a spike before undergoing small oscillations and settling to the late time post-revirialization state. This late time value is slightly lower (higher) than the one immediately after the encounter (marked by the spike) for higher (lower) encounter velocities. For more distant encounters, i.e., the case with $b=10\,r_\rmS$, $f_{\rm unbound}$ evolves more smoothly with time as the encounter only peels off particles from the outer shells of the haloes. In this case, the late time value of $f_{\rm unbound}$ is nearly the same as that shortly after the encounter (note the absence of a temporal spike in $f_{\rm unbound}$ unlike in the $b=r_\rmS$ case).

For each simulation, we compute the unbound fraction at the end of the simulation. This is roughly $150 \, t_{\rm dyn}$ after the encounter, where $t_{\rm dyn} = \sqrt{\frac{4}{3} \pi r^3_\rms/G M_\rms}$ is a characteristic dynamical time of the Hernquist sphere. We correct for the initial unbinding of $0.1\%$ of the particles by computing the stripped fraction, $f_{\rm strip} \equiv (f_{\rm unbound}-0.001)/0.999$. The solid triangles in the bottom panels of Fig.~\ref{fig:mass_loss} plot the resulting $f_{\rm strip}$ as a function of $v_{\rm P,peri}/\sigma$, where $v_{\rm P,peri}$ is the encounter velocity at pericenter. Due to gravitational focusing $v_{\rm P,peri}$ is somewhat larger than $v_\rmP$. For the $b=r_\rmS$ case, we also compute $f_{\rm strip}$ immediately after the encounter, and thus prior to re-virialization. These values are indicated by the solid circles in the bottom left panel.

As expected, encounters of higher velocity result in less mass loss. For the strongly penetrating encounters with $b=r_\rmS$, the encounter unbinds as much as $31$\% of the mass for the smallest encounter velocity considered here, $\vp = 0.7\,\sigma$ ($v_{\rm P,peri}=1.2\,\sigma$). For smaller encounter velocities, the two haloes actually become bound, resulting in tidal capture and ultimately a merger (indicated by the grey shaded region). For $\vp = \sigma$ ($v_{\rm P,peri}=1.4\,\sigma$), the stripped mass fraction is only $\sim 15$\%.

In the case of the larger impact parameter, $b=10\,r_\rmS$, tidal capture requires somewhat lower encounter speeds ($\vp \lta 0.35\,\sigma$, i.e., $v_{\rm {P,peri}} \lta 0.8\,\sigma$), and overall the encounter is significantly less damaging than for $b=r_\rmS$, with $f_{\rm strip} \sim 6$\% for $\vp = \sigma$ ($v_{\rm P,peri}=1.15\,\sigma$). For encounter speeds a little larger than the tidal capture value, about $25\%$ of the mass is stripped. Hence, we can conclude that penetrating hyperbolic encounters between two identical Hernquist spheres can result in appreciable mass loss ($\sim 25-30$\%), but only for encounter velocities that are close to the critical velocity that results in tidal capture. For any $v_\rmP > 2\,\sigma$ ($4\,\sigma$), the stripped mass fraction is less than 5\% (1\%), even for a strongly penetrating encounter. We therefore conclude that impulsive encounters between highly concentrated, cuspy systems, such as Hernquist or NFW spheres, rarely cause a significant mass loss.

\subsection{Comparison with predictions from our formalism}

We now turn to our generalized formalism in order to predict $f_{\rm strip}$ for the different encounters simulated above. We assume that the two haloes encounter each other along a straight-line orbit with impact parameter $b_{\rm peri}$ and encounter speed $v_{\rm P,peri}$. These values are inferred from the simulation results, and differ from the initial $b$ and $v_\rmP$ due to gravitational focusing. We then compute the impulse $\Delta \bv(\br)$ for each subject star using equation~(\ref{deltav1}). In the impulsive limit ($\vp \gg \sigma$), the encounter imparts to a single subject star a specific internal energy given by
\begin{align}
\Delta \varepsilon(\br) = \bv \cdot \Delta \bv_{\rm rel} + \frac{1}{2} {\left(\Delta v_{\rm rel}\right)}^2\,,
\label{dEstarrel}
\end{align}
where $\Delta \bv_{\rm rel}(\br) = \Delta \bv(\br) - \Delta \bv_{\rm CM}$. Using our formalism for a straight-line orbit, $\Delta \bv_{\rm CM}$ is given by equation~(\ref{deltavCM}).

To compute the fraction of subject stars that become unbound, $f_{\rm strip}$, we use the Monte Carlo method of \cite{vdBosch.etal.18a} and sample the isotropic equilibrium distribution function for a Hernquist sphere with $10^6$ particles each. We then follow two different methods of calculating $f_{\rm strip}$. 

In the first method, we consider a subject star to be stripped if its $\Delta \varepsilon/|\varepsilon| > 1$, where $\varepsilon =v^2/2 +\Phi_\rmS$ is the original binding energy of the star prior to the encounter. This equates $f_{\rm strip}$ to the fraction of particles that are deemed unbound immediately following the encounter, in the COM frame of the subject\footnote{The COM velocity is computed using all particles, both bound and unbound.}. The solid green lines in the bottom panels of Fig.~\ref{fig:mass_loss} plot the $f_{\rm strip}$ thus obtained as a function of $v_{\rm P,peri}/\sigma$.

In the second method, we compute $f_{\rm strip}$ in an iterative fashion. This is motivated by \citet[][hereafter AW85]{Aguilar.White.85}, who argued that the maximal subset of self-bound particles is a better predictor of the stripped mass fraction after revirialization. In the first iteration we simply calculate the number of stars that remain bound according to the criterion of the first method. Next we compute the center of mass position and velocity from only the bound particles identified in the previous iteration, which we use to recompute the impulse $\Delta \bv_{\rm rel}(\br)$. We also recompute the potential $\Phi_\rmS$ by constructing trees comprising of only these bound particles. Now we recalculate the number of bound particles using the new $\Delta \bv_{\rm rel}(\br)$ and $\Phi_\rmS$. We perform these iterations until the bound fraction and the center of mass position and velocity of the haloes converge. The $f_{\rm strip}$ thus obtained is indicated by the dashed green lines in Fig.~\ref{fig:mass_loss}. 

Overall, both methods yield stripped mass fractions that are in good agreement with each other and with the simulation results. Only when $v_{\rm P,peri}$ is close to the critical velocity for tidal capture does the iterative method yield somewhat larger $f_{\rm strip}$ than without iteration. For the $b=r_\rmS$ case, the Monte Carlo predictions agree well with the simulation results shortly after the encounter (solid circles), while slightly overestimating (underestimating) the post-revirialization $f_{\rm strip}$ values (solid triangles) for high (low) $v_{\rm P,peri}$. This is expected since the Monte Carlo formalism based on the impulse approximation does not account for the unbinding or rebinding of material due to revirialization. We find that the iterative approach to compute the maximal subset of self-bound particles does not significantly improve this. For the $b=10\,r_\rmS$ case, revirialization has very little impact and the Monte Carlo predictions match the simulation results very well. 

We have also repeated this exercise using the initial encounter speed and impact parameter, i.e., ignoring the impact of gravitational focusing. The results (not shown) are largely indistinguishable from those shown by the green curves, except at the low velocity end ($v_{\rm P,peri}/\sigma \lta 1$) of the $b=10\,r_\rmS$ case where it underestimates $f_{\rm strip}$. In the strongly penetrating case ($b=r_\rmS$) the effect of gravitational focusing is much weaker because the impulse has a reduced dependence on the impact parameter. Hence, gravitational focusing is only important if both $v_\rmp \lta \sigma$ and $b \gta r_\rmS$. Finally, we have also investigated the impact of adiabatic correction, which we find to have a negligible impact on $f_{\rm strip}$, unless the encounter (almost) results in tidal capture. 

The magenta, dashed lines in the lower panels of Fig.~(\ref{fig:mass_loss}) show the predictions for the stripped mass fractions provided by the fitting function of AW85. These were obtained by calculating the maximal subset of self-bound particles using a similar Monte Carlo method as used here, and using the fully general, non-perturbative expression for the impulse (their equation [3], which is equivalent to our equation~[\ref{deltav1}]). Although this fitting function matches well with our iterative formalism (green dashed lines) at the low velocity end, it significantly underestimates the stripped fractions for large $v_{\rm P,peri}/\sigma$. For these encounter velocities, the impulse is small and therefore unbinds only the particles towards the outer part of the halo (those with small escape velocities). The AW85 fitting function is based on encounters between two identical, $r^{1/4}$ \cite{deVaucouleurs.1948} spheres. These have density profiles that fall-off exponentially at large radii, and are thus far less `extended' than the Hernquist spheres considered here. Hence, it should not come as a surprise that their fitting function is unable to accurately describe the outcome of our experiments.

Finally, for comparison, the red lines in the bottom panels of Fig.~\ref{fig:mass_loss} correspond to the $f_{\rm strip}$ predicted using the DTA of GHO99. Here we again use the Monte-Carlo method, but the impulse for each star is computed using equation~(10) of GHO99 for the impact parameter, $b_{\rm peri}$, and encounter velocity, $v_{\rm P,peri}$ at pericentre (i.e., accounting for gravitational focusing). Although the DTA is clearly not valid for penetrating encounters, we merely show it here to emphasize that pushing the DTA into a regime where it is not valid can result in large errors. Even for impact parameters as large as $10\,r_\rmS$, the DTA drastically overpredicts $f_{\rm strip}$, especially at the high velocity end. High velocity encounters only strip off particles from the outer shells, for which the DTA severely overestimates the impulse. This highlights the merit of the general formalism presented here, which remains valid in those parts of the parameter space where the DTA breaks down.

To summarize, despite several simplifications such as the assumption of a straight line orbit, the impulse approximation, and the neglect of re-virialization, the generalized formalism presented here can be used to make reasonably accurate predictions for the amount of mass stripped off due to gravitational encounters between collisionless systems, even if the impact parameter is small compared to the characteristic sizes of the objects involved. In particular, in agreement with AW85, we find that the impulse approximation remains reasonably accurate all the way down to encounters that almost result in tidal capture, and which are thus no longer strictly impulsive.

\section{Conclusions}
\label{sec:conclusion}

In this paper we have developed a general, non-perturbative formalism to compute the energy transferred due to an impulsive shock. Previous studies \citep[e.g.,][]{Spitzer.58, Ostriker.etal.72, Richstone.75, Mamon.92, Mamon.00, Makino.Hut.97, Gnedin.etal.99} have all treated impulsive encounters in the distant tide limit by expanding the perturber potential as a multipole series truncated at the quadrupole term. However, this typically only yields accurate results if the impact parameter, $b$, is significantly larger than the characteristic sizes of both the subject, $r_\rmS,$ and the perturber, $r_\rmP$.  For such distant encounters, though, very little energy is transferred to the subject and such cases are therefore of limiting astrophysical interest. A noteworthy exception is the case where $r_\rmP \gg r_\rmS$, for which the formalism of GHO99, which also relies on the DTA, yields accurate results even when $b \ll r_\rmP$. However, even in this case, the formalism fails for impact parameters that are comparable to, or smaller than, the size of the subject. 

From an astrophysical perspective, the most important impulsive encounters are those for which the increase in internal energy, $\Delta E_{\rm int}$, is of order the subject's internal binding energy or larger. Such encounters can unbind large amounts of mass from the subject, or even completely destroy it. Unfortunately, this typically requires small impact parameters for which the DTA is no longer valid. In particular, when the perturber is close to the subject, the contribution of higher-order multipole moments of the perturber potential can no longer be neglected. The non-perturbative method presented here \citep[and previously in][]{Aguilar.White.85} overcomes these problems, yielding a method to accurately compute the velocity impulse on a particle due to a high-speed gravitational encounter. It can be used to reliably compute the internal energy change of a subject that is valid for any impact parameter, and any perturber profile. And although the results presented here are, for simplicity, limited to spherically symmetric perturbers, it is quite straightforward to extend it to axisymmetric, spheroidal perturbers, which is something we leave for future work.

In general, our treatment yields results that are in excellent agreement with those obtained using the DTA, but only if (i) the impact parameter  $b$ is large compared to the characteristic radii of both the subject and the perturber, and (ii) the subject is truncated at a radius $r_{\rm trunc} < b$.  If these conditions are not met, the DTA typically drastically overpredicts $\Delta E_{\rm int}$, unless one `manually' caps  $\Delta E_{\rm int}(b)$ to be no larger than the value for a head-on encounter, $\Delta E_0$ \citep[see e.g.,][]{vdBosch.etal.18a}. The $\Delta E_{\rm int}(b)$ computed using our fully general, non-perturbative formalism presented here, on the other hand, naturally asymptotes towards $\Delta E_0$ in the limit $b \to 0$. Moreover, in the DTA, a radial truncation of the subject is required in order to avoid divergence of the moment of inertia, $\langle r^2 \rangle$. Our method has the additional advantage that it does not suffer from this divergence-problem.

Although our formalism is more general than previous formalisms, it involves a more demanding numerical computation. In order to facilitate the use of our formalism, we have provided a table with the integrals $I(s)$ needed to compute the velocity impulse, $\Delta \bv(\br)$, given by equation~(\ref{deltav1}), for a variety of perturber profiles (Table~\ref{tab:Is}). In addition, we have released a public Python code, {\tt NP-impulse} (\url{https://github.com/uddipanb/NP-impulse}) that the reader can use to compute $\Delta \bv(\br)$ of a subject star as a function of impact parameter, $b$, and encounter speed, $\vp$. {\tt NP-impulse} also computes the resulting $\Delta E_{\rm int}$ for a variety of spherical subject profiles, and treats both straight-line orbits as well as eccentric orbits within the extended potential of a spherical perturber. In the latter case, {\tt NP-impulse} accounts for adiabatic shielding using the method developed in \cite{Gnedin.Ostriker.99}. We hope that this helps to promote the use of our formalism in future treatments of impulsive encounters.

As an example astrophysical application of our formalism, we have studied the mass loss experienced by a Hernquist sphere due to the tidal shock associated with an impulsive encounter with an identical object along a straight-line orbit. In general, our more general formalism agrees well with the results from numerical simulations and predicts that impulsive encounters are less disruptive, i.e., cause less mass loss, than what is predicted based on the DTA of GHO99. Encounters with $\vp/\sigma > 1$ do not cause any significant mass loss ($\lta 15$\%). For smaller encounter speeds, mass loss can be appreciable (up to $\sim 30$\%), especially for smaller impact parameters. However, for too low encounter speeds, $\vp/\sigma \lta 0.5$, the encounter results in tidal capture, and eventually a merger, something that cannot be treated using the impulse approximation. In addition, for $\vp/\sigma \lta 1$, the adiabatic correction starts to become important. Unfortunately, the adiabatic correction of \cite{Gnedin.Ostriker.99} that we have adopted in this paper has only been properly tested for the case of disc shocking, which involves fully compressive tides. It remains to be seen whether it is equally valid for the extensive tides considered here. Ultimately, in this regime a time-dependent perturbation analysis similar to that developed in \citet{Weinberg.94b} may be required to accurately treat the impact of gravitational shocking. Hence, whereas our formalism is fully general in the truly impulsive regime, and for any impact parameter, the case of slow, non-impulsive encounters requires continued, analytical studies. A particularly interesting case to examine is the quasi-resonant tidal interaction between a disk galaxy and a satellite. This has been explored in detail by \citet{DOnghia.etal.10}, who computed the impulse on disk stars while accounting for their rotation in the disk, rather than treating them as stationary. The impulse, however, was obtained perturbatively, using the DTA, and it remains to be seen how these results change if the impulse is computed non-perturbatively, as advocated here. We intend to address this in future work.

\section*{Acknowledgements}

The authors are grateful to Oleg Gnedin, Jerry Ostriker and the anonymous referee for insightful comments, and to Dhruba Dutta-Chowdhury and Nir Mandelker for valuable discussions. Special thanks are due to Simon White for pointing out our initial, unintentional oversight of Aguilar \& White (1985), who were the first to explicitly write out an expression for the impulse in the fully general case, and for his relentless vigilance that exposed a significant error in an earlier version of this paper. FvdB is supported by the National Aeronautics and Space Administration through Grant Nos. 17-ATP17-0028 and 19-ATP19-0059 issued as part of the Astrophysics Theory Program, and received additional support from the Klaus Tschira foundation.

\section*{Data availability}

The data underlying this article, including the Python code {\tt NP-impulse}, is publicly available in the GitHub Repository, at \url{https://github.com/uddipanb/NP-impulse}.


\bibliographystyle{mnras}
\bibliography{references_vdb} 

\begin{thebibliography}{}
\makeatletter
\relax
\def\mn@urlcharsother{\let\do\@makeother \do\$\do\&\do\#\do\^\do\_\do\%\do\~}
\def\mn@doi{\begingroup\mn@urlcharsother \@ifnextchar [ {\mn@doi@}
  {\mn@doi@[]}}
\def\mn@doi@[#1]#2{\def\@tempa{#1}\ifx\@tempa\@empty \href
  {http://dx.doi.org/#2} {doi:#2}\else \href {http://dx.doi.org/#2} {#1}\fi
  \endgroup}
\def\mn@eprint#1#2{\mn@eprint@#1:#2::\@nil}
\def\mn@eprint@arXiv#1{\href {http://arxiv.org/abs/#1} {{\tt arXiv:#1}}}
\def\mn@eprint@dblp#1{\href {http://dblp.uni-trier.de/rec/bibtex/#1.xml}
  {dblp:#1}}
\def\mn@eprint@#1:#2:#3:#4\@nil{\def\@tempa {#1}\def\@tempb {#2}\def\@tempc
  {#3}\ifx \@tempc \@empty \let \@tempc \@tempb \let \@tempb \@tempa \fi \ifx
  \@tempb \@empty \def\@tempb {arXiv}\fi \@ifundefined
  {mn@eprint@\@tempb}{\@tempb:\@tempc}{\expandafter \expandafter \csname
  mn@eprint@\@tempb\endcsname \expandafter{\@tempc}}}

\bibitem[\protect\citeauthoryear{{Aguilar} \& {White}}{{Aguilar} \&
  {White}}{1985}]{Aguilar.White.85}
{Aguilar} L.~A.,  {White} S.~D.~M.,  1985, \mn@doi [\apj] {10.1086/163382},
  \href {http://adsabs.harvard.edu/abs/1985ApJ...295..374A} {295, 374}

\bibitem[\protect\citeauthoryear{{Bahcall}, {Hut}  \& {Tremaine}}{{Bahcall}
  et~al.}{1985}]{Bahcall.etal.85}
{Bahcall} J.~N.,  {Hut} P.,   {Tremaine} S.,  1985, \mn@doi [\apj]
  {10.1086/162953}, \href
  {https://ui.adsabs.harvard.edu/abs/1985ApJ...290...15B} {290, 15}

\bibitem[\protect\citeauthoryear{{Barnes} \& {Hut}}{{Barnes} \&
  {Hut}}{1986}]{Barnes.Hut.86}
{Barnes} J.,  {Hut} P.,  1986, \mn@doi [\nat] {10.1038/324446a0}, \href
  {http://adsabs.harvard.edu/abs/1986Natur.324..446B} {324, 446}

\bibitem[\protect\citeauthoryear{{Binney}}{{Binney}}{2014}]{Binney.14}
{Binney} J.,  2014, arXiv e-prints, \href
  {https://ui.adsabs.harvard.edu/abs/2014arXiv1411.4937B} {p. arXiv:1411.4937}

\bibitem[\protect\citeauthoryear{{Binney} \& {Tremaine}}{{Binney} \&
  {Tremaine}}{1987}]{Binney.Tremaine.87}
{Binney} J.,  {Tremaine} S.,  1987, {Galactic dynamics}.
Princeton University Press

\bibitem[\protect\citeauthoryear{{Cappellari}}{{Cappellari}}{2002}]{Cappellari.02}
{Cappellari} M.,  2002, \mn@doi [\mnras] {10.1046/j.1365-8711.2002.05412.x},
  \href {https://ui.adsabs.harvard.edu/abs/2002MNRAS.333..400C} {333, 400}

\bibitem[\protect\citeauthoryear{{D'Onghia}, {Vogelsberger}, {Faucher-Giguere}
  \& {Hernquist}}{{D'Onghia} et~al.}{2010}]{DOnghia.etal.10}
{D'Onghia} E.,  {Vogelsberger} M.,  {Faucher-Giguere} C.-A.,   {Hernquist} L.,
  2010, \mn@doi [\apj] {10.1088/0004-637X/725/1/353}, \href
  {https://ui.adsabs.harvard.edu/abs/2010ApJ...725..353D} {725, 353}

\bibitem[\protect\citeauthoryear{{Dutta Chowdhury}, {van den Bosch}  \& {van
  Dokkum}}{{Dutta Chowdhury} et~al.}{2020}]{DuttaChowdhury.etal.20}
{Dutta Chowdhury} D.,  {van den Bosch} F.~C.,   {van Dokkum} P.,  2020, \mn@doi
  [\apj] {10.3847/1538-4357/abb947}, \href
  {https://ui.adsabs.harvard.edu/abs/2020ApJ...903..149D} {903, 149}

\bibitem[\protect\citeauthoryear{{Emsellem}, {Monnet}  \& {Bacon}}{{Emsellem}
  et~al.}{1994}]{Emsellem.etal.94}
{Emsellem} E.,  {Monnet} G.,   {Bacon} R.,  1994, \aap, \href
  {https://ui.adsabs.harvard.edu/abs/1994A&A...285..723E} {285, 723}

\bibitem[\protect\citeauthoryear{{Fabian}, {Pringle}  \& {Rees}}{{Fabian}
  et~al.}{1975}]{Fabian.etal.75}
{Fabian} A.~C.,  {Pringle} J.~E.,   {Rees} M.~J.,  1975, \mn@doi [\mnras]
  {10.1093/mnras/172.1.15P}, \href
  {https://ui.adsabs.harvard.edu/abs/1975MNRAS.172P..15F} {172, 15}

\bibitem[\protect\citeauthoryear{{Gnedin} \& {Ostriker}}{{Gnedin} \&
  {Ostriker}}{1999}]{Gnedin.Ostriker.99}
{Gnedin} O.~Y.,  {Ostriker} J.~P.,  1999, \mn@doi [\apj] {10.1086/306864},
  \href {http://adsabs.harvard.edu/abs/1999ApJ...513..626G} {513, 626}

\bibitem[\protect\citeauthoryear{{Gnedin}, {Hernquist}  \& {Ostriker}}{{Gnedin}
  et~al.}{1999}]{Gnedin.etal.99}
{Gnedin} O.~Y.,  {Hernquist} L.,   {Ostriker} J.~P.,  1999, \mn@doi [\apj]
  {10.1086/306910}, \href {http://adsabs.harvard.edu/abs/1999ApJ...514..109G}
  {514, 109}

\bibitem[\protect\citeauthoryear{{Heggie}}{{Heggie}}{1975}]{Heggie.75}
{Heggie} D.~C.,  1975, \mn@doi [\mnras] {10.1093/mnras/173.3.729}, \href
  {https://ui.adsabs.harvard.edu/abs/1975MNRAS.173..729H} {173, 729}

\bibitem[\protect\citeauthoryear{{Henon}}{{Henon}}{1959}]{Henon.59}
{Henon} M.,  1959, Annales d'Astrophysique, \href
  {https://ui.adsabs.harvard.edu/abs/1959AnAp...22..126H} {22, 126}

\bibitem[\protect\citeauthoryear{{Hernquist}}{{Hernquist}}{1990}]{Hernquist.90}
{Hernquist} L.,  1990, \mn@doi [\apj] {10.1086/168845}, \href
  {http://adsabs.harvard.edu/abs/1990ApJ...356..359H} {356, 359}

\bibitem[\protect\citeauthoryear{{Lee} \& {Ostriker}}{{Lee} \&
  {Ostriker}}{1986}]{Lee.Ostriker.86}
{Lee} H.~M.,  {Ostriker} J.~P.,  1986, \mn@doi [\apj] {10.1086/164674}, \href
  {https://ui.adsabs.harvard.edu/abs/1986ApJ...310..176L} {310, 176}

\bibitem[\protect\citeauthoryear{{Makino} \& {Hut}}{{Makino} \&
  {Hut}}{1997}]{Makino.Hut.97}
{Makino} J.,  {Hut} P.,  1997, \mn@doi [\apj] {10.1086/304013}, \href
  {https://ui.adsabs.harvard.edu/abs/1997ApJ...481...83M} {481, 83}

\bibitem[\protect\citeauthoryear{{Mamon}}{{Mamon}}{1992}]{Mamon.92}
{Mamon} G.~A.,  1992, \mn@doi [\apjl] {10.1086/186656}, \href
  {https://ui.adsabs.harvard.edu/abs/1992ApJ...401L...3M} {401, L3}

\bibitem[\protect\citeauthoryear{{Mamon}}{{Mamon}}{2000}]{Mamon.00}
{Mamon} G.~A.,  2000, in {Combes} F.,  {Mamon} G.~A.,   {Charmandaris} V.,
  eds,  Astronomical Society of the Pacific Conference Series Vol. 197,
  Dynamics of Galaxies: from the Early Universe to the Present. p.~377
  (\mn@eprint {arXiv} {astro-ph/9911333})

\bibitem[\protect\citeauthoryear{{Martinez-Medina}, {Gieles}, {Gnedin}  \&
  {Li}}{{Martinez-Medina} et~al.}{2020}]{Martinez-Medina.etal.20}
{Martinez-Medina} L.~A.,  {Gieles} M.,  {Gnedin} O.~Y.,   {Li} H.,  2020, arXiv
  e-prints, \href {https://ui.adsabs.harvard.edu/abs/2020arXiv200906643M} {p.
  arXiv:2009.06643}

\bibitem[\protect\citeauthoryear{{Moore}, {Katz}, {Lake}, {Dressler}  \&
  {Oemler}}{{Moore} et~al.}{1996}]{Moore.etal.96b}
{Moore} B.,  {Katz} N.,  {Lake} G.,  {Dressler} A.,   {Oemler} A.,  1996,
  \mn@doi [\nat] {10.1038/379613a0}, \href
  {http://adsabs.harvard.edu/abs/1996Natur.379..613M} {379, 613}

\bibitem[\protect\citeauthoryear{{Navarro}, {Frenk}  \& {White}}{{Navarro}
  et~al.}{1997}]{Navarro.etal.97}
{Navarro} J.~F.,  {Frenk} C.~S.,   {White} S.~D.~M.,  1997, \mn@doi [\apj]
  {10.1086/304888}, \href {http://adsabs.harvard.edu/abs/1997ApJ...490..493N}
  {490, 493}

\bibitem[\protect\citeauthoryear{{Ostriker}, {Spitzer}  \&
  {Chevalier}}{{Ostriker} et~al.}{1972}]{Ostriker.etal.72}
{Ostriker} J.~P.,  {Spitzer} Lyman J.,   {Chevalier} R.~A.,  1972, \mn@doi
  [\apjl] {10.1086/181018}, \href
  {https://ui.adsabs.harvard.edu/abs/1972ApJ...176L..51O} {176, L51}

\bibitem[\protect\citeauthoryear{{Plummer}}{{Plummer}}{1911}]{Plummer.11}
{Plummer} H.~C.,  1911, \mn@doi [\mnras] {10.1093/mnras/71.5.460}, \href
  {https://ui.adsabs.harvard.edu/abs/1911MNRAS..71..460P} {71, 460}

\bibitem[\protect\citeauthoryear{{Press} \& {Teukolsky}}{{Press} \&
  {Teukolsky}}{1977}]{Press.Teukolsky.77}
{Press} W.~H.,  {Teukolsky} S.~A.,  1977, \mn@doi [\apj] {10.1086/155143},
  \href {https://ui.adsabs.harvard.edu/abs/1977ApJ...213..183P} {213, 183}

\bibitem[\protect\citeauthoryear{{Richstone}}{{Richstone}}{1975}]{Richstone.75}
{Richstone} D.~O.,  1975, \mn@doi [\apj] {10.1086/153820}, \href
  {https://ui.adsabs.harvard.edu/abs/1975ApJ...200..535R} {200, 535}

\bibitem[\protect\citeauthoryear{{Richstone}}{{Richstone}}{1976}]{Richstone.76}
{Richstone} D.~O.,  1976, \mn@doi [\apj] {10.1086/154213}, \href
  {https://ui.adsabs.harvard.edu/abs/1976ApJ...204..642R} {204, 642}

\bibitem[\protect\citeauthoryear{{Spitzer}}{{Spitzer}}{1958}]{Spitzer.58}
{Spitzer} Jr. L.,  1958, \mn@doi [\apj] {10.1086/146435}, \href
  {http://adsabs.harvard.edu/abs/1958ApJ...127...17S} {127, 17}

\bibitem[\protect\citeauthoryear{{Weinberg}}{{Weinberg}}{1994a}]{Weinberg.94a}
{Weinberg} M.~D.,  1994a, \mn@doi [\aj] {10.1086/117161}, \href
  {http://adsabs.harvard.edu/abs/1994AJ....108.1398W} {108, 1398}

\bibitem[\protect\citeauthoryear{{Weinberg}}{{Weinberg}}{1994b}]{Weinberg.94b}
{Weinberg} M.~D.,  1994b, \mn@doi [\aj] {10.1086/117162}, \href
  {http://adsabs.harvard.edu/abs/1994AJ....108.1403W} {108, 1403}

\bibitem[\protect\citeauthoryear{{White}}{{White}}{1978}]{White.78}
{White} S.~D.~M.,  1978, \mn@doi [\mnras] {10.1093/mnras/184.2.185}, \href
  {https://ui.adsabs.harvard.edu/abs/1978MNRAS.184..185W} {184, 185}

\bibitem[\protect\citeauthoryear{{de Vaucouleurs}}{{de
  Vaucouleurs}}{1948}]{deVaucouleurs.1948}
{de Vaucouleurs} G.,  1948, Annales d'Astrophysique, \href
  {https://ui.adsabs.harvard.edu/abs/1948AnAp...11..247D} {11, 247}

\bibitem[\protect\citeauthoryear{{van den Bosch} \& {Ogiya}}{{van den Bosch} \&
  {Ogiya}}{2018}]{vdBosch.etal.18b}
{van den Bosch} F.~C.,  {Ogiya} G.,  2018, \mn@doi [\mnras]
  {10.1093/mnras/sty084}, \href
  {http://adsabs.harvard.edu/abs/2018MNRAS.475.4066V} {475, 4066}

\bibitem[\protect\citeauthoryear{{van den Bosch}, {Ogiya}, {Hahn}  \&
  {Burkert}}{{van den Bosch} et~al.}{2018}]{vdBosch.etal.18a}
{van den Bosch} F.~C.,  {Ogiya} G.,  {Hahn} O.,   {Burkert} A.,  2018, \mn@doi
  [\mnras] {10.1093/mnras/stx2956}, \href
  {http://adsabs.harvard.edu/abs/2018MNRAS.474.3043V} {474, 3043}

\makeatother
\end{thebibliography}


\appendix

\section{Asymptotic behaviour}
\label{app:asymptote}

In \S\ref{sec:straight_orbit}, we obtained the general expression for $\Delta E_{\rm int}$, which is valid for impulsive encounters with any impact parameter $b$. Here we discuss the asymptotic behaviour of $\Delta E_{\rm int}$ in both the distant tide limit (large $b$) and the head-on limit (small $b$).

\subsection{Distant encounter approximation}
\label{sec:asymptote_tidal}

In the limit of distant encounters, the impact parameter $b$ is much larger than the scale radii of the subject, $r_\rmS$, and the perturber, $r_\rmP$. In this limit, it is common to approximate the perturber as a point mass. However, as discussed above, this will yield a diverging $\Delta E_{\rm int}$ unless the subject is truncated and $b > r_{\rm trunc}$ (an assumption that is implied, but rarely mentioned). In order to avoid this issue, we instead consider a (spherical) Plummer perturber. In the limit of large $b$,  equation~(\ref{delEint1}) then reduces to an expression that is similar to, but also intriguingly different from, the standard distant tide expression first obtained by S58 by treating the perturber as a point mass, and expanding $\Phi_\rmP$ as a multipole series truncated at the quadrupole term. We also demonstrate that the asymptotic form of $\Delta E_{\rm int}$ is quite different for infinite and truncated subjects.

In the large-$b$ limit, we can assume that $r\sin{\theta} < b$, i.e., we can restrict the domains of the $\calJ$ and $\calJ_{\rm CM}$ integrals (equations~[\ref{J_plummer}] and~[\ref{J_CM_plummer}]) to the inside of a cylinder of radius $b$. The use of cylindrical coordinates is prompted by the fact that the problem is inherently cylindrical in nature: the impulse received by a subject star is independent of its distance along the direction in which the perturber is moving, but only depends on $R = r \sin\theta$ (cf. equation~[\ref{deltav0}]).  Hence, in computing the total energy change, $\Delta E$, it is important to include subject stars with small $R$ but large $z$-component, while, in the DTA, those with $R > b$ can be ignored as they receive a negligibly small impulse. Next, we Taylor expand the $\theta$-integrand in the expression for $\calJ$ about $r\sin{\theta}=0$ to obtain the following series expansion for the total energy change
\begin{align}
&\Delta E \approx 4\pi{\left(\frac{GM_\rmP}{\vp}\right)}^2 \int_0^{\infty} \rmd r\, r^2 \rho_\rmS(r) \int_0^{\pi} \rmd \theta \sin{\theta} \left[\frac{1}{{\left(1+\varepsilon^2\right)}^2}\frac{1}{b^2}+\frac{1-4\varepsilon^2+\varepsilon^4}{{\left(1+\varepsilon^2\right)}^4}\frac{r^2\sin^2{\theta}}{b^4}+\frac{1-12\varepsilon^2+15\varepsilon^4-2\varepsilon^6}{{\left(1+\varepsilon^2\right)}^6}\frac{r^4\sin^4{\theta}}{b^6}+...\right]\,,
\label{deltaEint_series}
\end{align}
where $\varepsilon=r_\rmP/b$. In the large $b$ limit, the COM velocity given by equation~(\ref{deltavCM_ptmass}) reduces to
\begin{align}
\Delta \bv_{\rm CM} = \frac{2GM_\rmP}{M_\rmS\vp}\frac{\pi}{b}\int_0^{\infty} \rmd r\, r^2 \rho_\rmS(r) \int_0^{\pi} \rmd \theta \sin{\theta}\left[\frac{2}{1+\varepsilon^2}-\frac{4\varepsilon^2}{{\left(1+\varepsilon^2\right)}^3}\frac{r^2\sin^2{\theta}}{b^2}+...\right]\, \hat{\by}\,.
\end{align}
The above two integrals have to be evaluated conditional to $r\sin{\theta}<b$. Upon subtracting the COM energy, $\Delta E_{\rm CM} = \frac{1}{2} M_\rmS \, (\Delta v_{\rm CM})^2$, the first term in the $\theta$ integrand of equation~(\ref{deltaEint_series}) drops out. Integrating the remaining terms yields
\begin{align}
&\Delta E_{\rm int} \approx 4\pi {\left(\frac{GM_\rmP}{\vp}\right)}^2 \sum_{n=2}^{\infty}\calI_{n-1}\,\calC_n\left(\frac{r_\rmP}{b}\right)\frac{\calR_n(b)+\calS_n(b)}{b^{2n}}\,.
\label{tidal_general}
\end{align}
Here
\begin{align}
\calC_n(x) = \frac{P_{2n}(x)}{(1+x^2)^{2n}},
\end{align}
with $P_{2n}(x)$ a polynomial of degree $2n$. We have worked out the coefficients for $n=2$ and $3$, yielding $P_4(x) = 1 + x^4$ and $P_6(x) = 1 - 6 x^2 + 9 x^4 - 2 x^6$, and leave the coefficients for the higher-order terms as an exercise for the reader. We do point out, though, that $\calC_n(r_\rmP/b) = 1+\calO(r^2_\rmP/b^2)$ in the limit $b \gg r_\rmP$. The coefficient $\calI_n$ is given by
\begin{align}
\calI_n & = \int_{-1}^1 \rmd x\, (1-x^2)^n = 2 \sum_{m=0}^{n} \, \frac{(-1)^m}{2m+1} \, \binom{n}{m} \,,
\end{align}
while $\calR_n(b)$ and $\calS_n(b)$ are functions of $b$ given by
\begin{align}
&\calR_n(b)=\int_0^b \rmd r\,r^{2n}\rho_\rmS(r)\,, \nonumber \\
&\calS_n(b)=\int_b^{\infty} \rmd r\,r^{2n} \rho_\rmS(r) \left[1-\sqrt{1-\frac{b^2}{r^2}}\left\{1+\sum_{m=0}^{n-2}\frac{\binom{2m+1}{m}}{2^{2m+1}}{\left(\frac{b}{r}\right)}^{2m+2}\right\}\right]\,.
\end{align}
Note that $\calR_n(b)$ is the ${\left(2n-2\right)}^{\rm th}$ moment of the subject density profile, $\rho_\rmS(r)$, inside a sphere of radius $b$, while $\calS_n(b)$ is the same but for the part of the cylinder outside of the sphere. $\calR_n(b)+\calS_n(b)$ is therefore the ${\left(2n-2\right)}^{\rm th}$ moment of $\rho_\rmS(r)$ within the cylinder of radius $b$. If we truncate the series given in equation~(\ref{tidal_general}) at $n=2$, then we obtain an asymptotic form for $\Delta E_{\rm int}$ that is similar to that of the standard tidal approximation:
\begin{align}
&\Delta E_{\rm int} \approx \frac{4 M_\rmS}{3} {\left(\frac{GM_\rmP}{\vp}\right)}^2 \frac{\langle r^2\rangle_{\rm cyl}}{b^4}\,.
\label{tidal}
\end{align}
Here
\begin{eqnarray}
\langle r^2\rangle_{\rm cyl} & = & \frac{4\pi}{M_\rmS} \left[\int_0^b \rmd r\, r^4 \rho_\rmS(r) + \int_b^{\infty} \rmd r\, r^4 \rho_\rmS(r)\left\{1-\sqrt{1-\frac{b^2}{r^2}}\left(1+\frac{b^2}{2r^2}\right)\right\} \right] \nonumber \\
&= & \langle r^2 \rangle - \frac{4\pi}{M_\rmS} \int_b^{\infty} \rmd r\, r^4 \rho_\rmS(r)\sqrt{1-\frac{b^2}{r^2}}\left(1+\frac{b^2}{2r^2}\right)\,.
\label{moment_of_inertia}
\end{eqnarray}
which is subtly different from the moment of inertia, $\langle r^2\rangle$, that appears in the standard expression for the distant tidal limit, and which is given by equation~(\ref{r2aver}). In particular, $\langle r^2 \rangle_{\rm cyl}$ only integrates the subject mass within a cylinder truncated at the impact parameter, whereas $\langle r^2 \rangle$ integrates over the entire subject mass. As discussed above, this typically results in a divergence, unless the subject is truncated or has a density that falls of faster than $r^{-5}$ in its outskirts.

Indeed, if the subject is truncated at a truncation radius $r_{\rm trunc} < b$, then $\langle r^2 \rangle_{\rm cyl} = \langle r^2 \rangle$, and equation~(\ref{tidal}) is exactly identical to that for the `standard' impulsive encounter of S58. In addition, $\calR_n=\int_0^{r_{\rm trunc}}\rmd r\,r^{2n}\rho_\rmS(r)$, which is independent of $b$, and $\calS_n=0$. Hence, the $n^{\rm th}$-order term scales as $b^{-2n}$, and $\Delta E_{\rm int}$ is thus dominated by the quadrupole term, justifying the truncation of the series in equation~(\ref{deltaEint_series}) at $n=2$.

However, for an infinitely extended subject, or one that is truncated at $r_{\rm trunc} > b$, truncating the series at the $n=2$ quadrupole term can, in certain cases, underestimate $\Delta E_{\rm int}$ by as much as a factor of $\sim 2$. In particular, if $\rho_\rmS(r) \sim r^{-\beta}$ at large $r$, and falls off less steeply than $r^{-5}$ at small $r$, then  both $\calR_n(b)$ and $\calS_n(b)$ scale as $b^{2n+1-\beta}$, as long as $\beta < 5$. Hence, all terms in equation~(\ref{tidal_general}) scale with $b$ in the same way, and the truncation is not justified, even in the limit of large impact parameters\footnote{This is also evident from equation~(\ref{deltaEint_series}), which shows that all terms contribute equally when  $r\sin{\theta}\sim b$.}. Furthermore, in this case it is evident from equation~(\ref{tidal_general}) that $\Delta E_{\rm int}\sim b^{1-\beta}$. On the other hand, for $\beta=5$, $\calR_2$ is the dominant term and scales with $b$ as $\ln{b}$, so that $\Delta E_{\rm int}\sim \ln{b}/b^4$. For $\beta>5$, both $\calR_2$ and $\calS_2$ are the dominant terms, which add up to $\langle r^2\rangle \simeq \int_0^{\infty}\rmd r\, r^4\rho_\rmS(r)$ (which is finite in this case), such that $\Delta E_{\rm int}\sim b^{-4}$. Hence, for an infinitely extended subject with $\rho_\rmS \propto r^{-\beta}$ at large $r$ we have that
\begin{equation}
\begin{aligned}
\lim_{b\to\infty}\Delta E_{\rm int} \propto
\begin{cases}
b^{1-\beta} , & \beta < 5 \\[2pt]
b^{-4} \ln{b}, & \beta = 5 \\[2pt]
b^{-4}, & \beta > 5\,.
\end{cases}
\label{asymptote_tidal_infinite}
\end{aligned}
\end{equation}
This scaling is not only valid for an infinitely extended subject, but also for a truncated subject when the impact parameter falls in the range $\max[r_\rmS,r_\rmP] < b < r_{\rm trunc}$.

\subsection{Head-on encounter approximation}
\label{sec:asymptote_head_on}

The head-on encounter corresponds to the case of zero impact parameter (i.e., $b=0$). As long as the perturber is not a point mass, the internal energy injected into the subject is finite, and can be computed using equation~(\ref{deltaE1}) with $b=0$. Note that there is no need to subtract $\Delta E_{\rm CM}$ in this case, since it is zero. If the perturber is a Plummer sphere, the $\calJ$ integral can be computed analytically for $b=0$, which yields
\begin{align}
\Delta E_{\rm int} = 8\pi{\left(\frac{G M_\rmP}{\vp}\right)}^2 \int_0^{\infty} \rmd r\,\rho_\rmS(r)\, \calF_0(r,r_\rmP),
\label{headon}
\end{align}
where
\begin{align}
\calF_0(r,r_\rmP) = \frac{r\left(2r^2+r^2_\rmP\right)}{4{\left(r^2+r^2_\rmP\right)}^{3/2}} \ln{\left[\frac{\sqrt{r^2+r^2_\rmP}+r}{\sqrt{r^2+r^2_\rmP}-r}\right]} - \frac{r^2}{2\left(r^2+r^2_\rmP\right)}.
\end{align}
It is easily checked that $\calF_0$ has the following asymptotic behaviour in the small- and large-$r$ limits:
\begin{equation}
\begin{aligned}
\calF_0(r,r_\rmP) \sim
\begin{cases}
 \frac{2}{3} \left(\frac{r}{r_\rmP}\right)^4, & r \ll r_\rmP \\[8pt]
 \ln{\left(\frac{2r}{r_\rmP}\right)}, & r \gg r_\rmP.
\end{cases}
\label{F0}
\end{aligned}
\end{equation}
Hence, we see that the behavior of the integrand of equation~(\ref{headon}) in the limits $r\to 0$ ($r\ll r_\rmP$) and $r\to \infty$ ($r\gg r_\rmP$), is such that $\Delta E_{\rm int}$ is finite, as long as $\rho_\rmS(r)$ scales less steeply than $r^{-5}$ at small $r$ and more steeply than $r^{-1}$ at large $r$. Both conditions are easily satisfied for any realistic astrophysical subject. Note from equation~(\ref{F0}) that, as expected, more compact perturbers (smaller $r_\rmP$) dissipate more energy and therefore cause more pronounced heating of the subject. 

Note that one obtains the same results using the expression of $\Delta E_{\rm int}$ for a head-on encounter listed under case C in Table~\ref{tab:comparison}. For a Plummer perturber, $I_0 = R^2/(R^2 + r^2_\rmp)$, which after substitution in the expression for $\Delta E_{\rm int}$, writing $R = r \sin\theta$, and solving the $\theta$-integral, yields equation~(\ref{headon}).

\bsp	
\label{lastpage}

\end{document}